# TrimMoE: A Communication-aware and Adaptive-depth Framework for Distributed Edge Inference

Ning Li, Shuting Bai, Xin Yuan, Wenchao Xu, Song Guo, *Fellow, IEEE*, Haijun Zhang, *Fellow, IEEE*

**Abstract—Serving Mixture-of-Experts (MoE) large language models across distributed edge servers is bottlenecked by the cross-server expert transmission. The existing approaches mainly focus on how to reach a remote expert faster. However, in this paper, we instead consider whether a given layer, and the layers after it, need to be executed at all. To this end, a communication-aware adaptive-depth framework is proposed in this paper, termed TrimMoE, which couples layer skipping and confidence-based early exit with substitute execution and server-expert selection under a unified quality budget. Specifically, in the offline stage, TrimMoE freezes the backbone, trains the lightweight per-layer exit heads, calibrates the per-layer importance thresholds, and allocates the expert replicas by a skip/exit-aware redundancy benefit. In the online stage, a transition-aware look-ahead anticipates the token movement, so that the depth reduction targets the costliest transmissions, and besides, two feedback rules adapt the delay-quality weights and the exit threshold. Moreover, we prove that the substitution-and-skipping proxy degradation never exceeds the configured budget, and that the early exit is admitted only under a calibrated confidence gate. On a heterogeneous 10-server testbed with Switch-Base-8E, Qwen-MoE-A2.7B, and Mixtral-8x7B, TrimMoE reduces the average latency by up to 62.8%, lowers the cross-server traffic and the remote-execution ratio, and sustains high throughput under load, while keeping the task-quality degradation within a 2% bound.**

## I. Introduction

Large language models (LLMs) have been widely used and have shown impressive performance in many intelligent applications, such as conversational agents, code generation, multimodal perception, interactive content creation, etc. [1]. However, because the parameter scale and inference cost of LLMs are always large, it is difficult to directly execute them on resource-limited end devices or individual edge servers [2,3]. To address this issue, edge intelligence has emerged as an appealing direction, in which the inference workloads are jointly carried out by terminals and nearby edge infrastructure to lower response time, save backbone bandwidth, and better protect user data. Among the different LLM frameworks, the mixture-of-experts (MoE) architecture is especially suitable for edge deployment, since each token only activates a small portion of the experts, which achieves a good balance between representational capacity and per-token computation [4,5].

To deploy LLMs on the edge, the existing works generally follow two routes. The first route is model compression, such as quantization, structured pruning [6,7], knowledge distillation [8], and low-rank factorization [9,10], which reduces the model footprint and computation cost before inference. However, model compression often reduces the inference accuracy, and for extremely large LLMs the compressed model can still exceed the memory and computation limits of a single edge node [11]. The second route is distributed inference, in which a large model is split and cooperatively executed over several servers [12]. Compared with model compression, distributed inference has better scalability, because the memory and computation can be amortized across multiple nodes [13]. Therefore, cooperative inference across edge servers becomes indispensable once a single server can neither hold the full model nor provide adequate inference throughput.

Distributed serving of large models has been extensively studied in cloud data centers, where the partitions of a model are executed over many servers to improve throughput [14,15]. In the cloud, the servers are usually connected by high speed transmission technologies, such as RDMA [16], InfiniBand [17], so the communication cost of cross-server execution can be well controlled. However, it is hard to directly apply these cloud-oriented designs to the edge. First, unlike the tightly coupled cloud, the geographically dispersed edge servers are usually connected over the public Internet rather than dedicated high-bandwidth links [18], so the inter-server bandwidth is much more limited, the latency is higher, and the time-varying network state further improves the complexity of this problem. Second, the edge servers differ widely in memory budget, computation capability, and load, which makes it more difficult to distribute LLMs across them [19].

These edge-specific properties make distributed MoE serving particularly difficult. In our previous work, the same-layer expert similarity is leveraged, so that a token whose routed

This paragraph of the first footnote will contain the date on which you submitted your paper for review, which is populated by IEEE. This work was supported in part by the grant from NSFC Grant no. 62571156, 62101159, 52475009, NSF of Shandong Grant no. ZR2021MF055, ZR2025QC666, the Research Grants Council of Hong Kong under the Areas of Excellence scheme grant AoE/E-601/22-R, and also the Opening Project of the Key Laboratory of Advanced Manufacturing and Intelligent Technology (Ministry of Education) at Harbin University of Science and Technology (KFKT202306).
*(Corresponding author: Xin Yuan).*

Ning Li and Haijun Zhang are with the School of Artificial Intelligence, University of Science and Technology Beijing, China (e-mail: ningli_polyuhk@outlook.com, zhanghaijun@ustb.edu.cn).

Xin Yuan is with the School of Ocean Engineering, Harbin Institute of Technology, Heilongjiang, China (e-mail: xin.yuan@hit.edu.cn).

Shuting Bai is with the School of Computer Science and Technology, Harbin Institute of Technology, Heilongjiang, China (e-mail: bstican@163.com).

Wenchao Xu is with the Division of Integrative Systems and Design, Hong Kong University of Science and Technology, Hong Kong (e-mail: wenchaoxu@ust.hk)

Song Guo is with the Department of Computer Science and Engineering, Hong Kong University of Science and Technology, Hong Kong (e-mail: songguo@cse.ust.hk)

Mentions of supplemental materials and animal/human rights statements can be included here.

Color versions of one or more of the figures in this article are available online at http://ieeexplore.ieee.org

expert is not available locally can be served by a functionally similar local substitute instead of being forwarded to a remote server, which already reduces a large fraction of cross-server transmissions [37]. However, there are still two kinds of inter-server traffic. First, every token is still pushed through all MoE layers, even though many layers contribute marginally to a given token, and a non-trivial part of these layers needs a remote hop to execute. Second, every token is still processed to the full depth, even when its representation has already stabilized at a much shallower layer. Most existing works concentrate on how to reach the remote computation unit more efficiently, but seldom consider whether it is necessary to traverse that layer, or the remaining layers, at all. Therefore, for distributed edge MoE inference, eliminating the unnecessary layer executions and the transmissions they trigger, rather than only accelerating them, is still an open and appealing problem.

Fortunately, recent works reveal that the computation of an MoE model is highly redundant along the depth for individual tokens [20]. First, the MoE layers are not equally important: the gating response of the routed expert provides a per-token and per-layer importance signal, and the layers with weak importance can be bypassed through the residual path with little impact on the final prediction [21]. Second, many tokens reach a representation that already matches the full-depth result well before the last layer, so a lightweight confidence estimator can decide that the remaining layers are unnecessary and terminate the inference early [22]. These two observations are particularly valuable in the distributed edge scenario. On the one hand, if a low-importance layer is one that would otherwise force a cross-server hop, skipping it removes the transmission entirely. On the other hand, if a token can confidently exit early, all the remaining transmission-prone layers are removed at once. In other words, layer skipping and early exit change the question from "how to transmit a token to the right remote expert" into "whether that layer, and the layers beyond it, need to be executed on the remote expert at all". To the best of our knowledge, this is the first work which jointly exploits communication-aware layer skipping and confidence-based early exit, coordinated with server-expert selection along the token trajectory, to further suppress the cross-server transmission in distributed edge MoE inference.

Nevertheless, it is far from trivial to turn this idea into a practical framework. First, the skip decision, the exit decision, and the server-expert choice are tightly coupled along the token's path, i.e., a skipped layer keeps the token on the current server while consuming part of the quality budget, an early exit cancels all the remaining layers once the confidence is met, and the server selected for an executed layer determines where the next layer sources the token. Therefore, handling these decisions in isolation is not enough. Second, skipping a layer or exiting early reduces the communication but may reduce the inference quality, so the framework must respect a controllable per-user quality degradation budget, while ensuring that the early exit never terminates on an under-refined representation. Third, a sound decision at the current layer depends on the transmission cost and the remaining budget of the layers that have not been computed yet, which needs a look-ahead mechanism to anticipate the future token movement without actually executing those layers. Finally, because the edge servers are heterogeneous in memory and computation, the offline deployment must concentrate the redundant expert replicas on the experts that genuinely reduce transmission after skipping and early exit are enabled, rather than on the experts that are frequently bypassed or rarely reached.

To address the above challenges, a communication-aware adaptive-depth framework for distributed edge MoE inference is proposed in this paper, termed TrimMoE. Specifically, in the offline stage, TrimMoE freezes the backbone and trains one lightweight per-layer exit head to produce a calibrated confidence score and calibrates a per-layer importance threshold so that only low-importance layers become skip candidates within a budget-derived tolerance. Then, it collects the expert transition statistics together with a compact token-similarity reference set for runtime prediction. Based on the similarity-aware deployment inherited from our previous work, TrimMoE further refines the redundant placement by a skip-exit-aware redundancy benefit, so that improving the efficiency of edge memory. In the online stage, for each token at each layer, TrimMoE jointly determines whether to skip, to exit, or to execute, and on which server with which expert. The purpose is to minimize the sum of a normalized per-step cost and a look-ahead cost-to-go estimated from the transition statistics. Besides, a pair of runtime feedback rules continuously adapts the delay-quality weights and the exit confidence threshold to the changing network and workload conditions. In this way, TrimMoE removes both the unnecessary layer executions and their induced transmissions over the bandwidth-limited and dynamic edge links, while keeping the inference quality within a controllable budget.

The main contributions of this paper can be summarized as follows.

- We formulate the distributed edge MoE inference problem as a joint optimization of layer skipping, early exit, and server-expert selection over the token trajectory. Different from the existing formulations that treat every layer as mandatory and every token as full-depth, the proposed formulation allows a token to skip the transmission-inducing low-importance layers and to terminate early under a confidence requirement, while updating the token location after each executed layer and respecting a per-user quality degradation budget.
- We design an offline calibration and communication-aware deployment scheme. Without fine-tuning the backbone, we train the per-layer exit heads on the cached representations to obtain a quality-preserving exit indicator, calibrate the layer importance thresholds from the empirical skip-degradation relationship, and refine the redundant expert deployment by a skip-exit-aware redundancy benefit, which prioritizes the experts whose replication genuinely reduces cross-server transmission.
- We develop a communication-aware online skip-exit and server-expert selection algorithm. The algorithm jointly weighs the transmission delay, the computation delay, the substitution and skip-induced quality loss, and the look-ahead cost of future token movement. It decides not only whether a layer should be skipped, executed, or exited, but also which server and expert should serve an executed layer, and it self-calibrates its weights and exit threshold from the runtime observations.
- We implement TrimMoE on a real distributed edge testbed.

The experimental results demonstrate that, compared with the representative baselines, TrimMoE markedly reduces the average latency, the tail latency, the cross-server traffic, and the remote execution ratio, while keeping the introduced inference quality degradation small and bounded by the configured budget.

The remainder of this paper is organized as follows. Section II surveys the related work. Section III describes the system model and formulates the optimization problem. Section IV presents the offline exit-head training and communication-aware deployment scheme. Section V details the communication-aware online skip-exit and server-expert selection algorithm. Section VI reports the experimental evaluation. Finally, Section VII concludes this paper.

## II. Related Works

In this section, we review the existing studies related to this paper from three aspects: 1) collaborative inference and distributed deployment of large models at the edge; 2) MoE serving optimization and communication reduction; and 3) layer skipping and early exit for efficient inference. Although these studies have improved the efficiency of large-model inference from different perspectives, they cannot effectively address the problem studied in this paper, because none of them jointly exploits communication-aware layer skipping and confidence-based early exit, coordinated with substitute-aware server-expert selection along the token trajectory, to remove the cross-server transmission triggered by the unnecessary layers in distributed edge MoE inference.

### *A. Collaborative Inference and Distributed Deployment of Large Models at the Edge*

The deployment and inference acceleration of large models in edge environments have been studied in many previous works. The early works mainly focus on model partitioning, collaborative inference, task offloading, and resource allocation, so as to reduce the end-to-end latency and improve the efficiency of resource-constrained edge intelligence. For conventional DNNs, Mohammed et al. [23] formulate the adaptive DNN partitioning and offloading as a matching game to accelerate the distributed inference, and PArtNNer [24] proposes a platform-agnostic adaptive edge-cloud DNN partitioning method, which automatically locates the optimal partition point to minimize the end-to-end latency. Besides, a comprehensive survey in [25] summarizes the split-computing and early-exiting techniques, together with their modeling and optimization objectives, for cooperative DNN inference at the edge.

With the rapid development of LLMs, the recent works have extended the collaborative inference from conventional DNNs to large generative models. Edge-LLM [26] proposes a collaborative framework for LLM serving in edge computing, and EdgeShard [12] partitions an LLM into shards and jointly optimizes the device selection and the model partition for collaborative edge inference. To further reduce the communication cost of tensor-parallel inference, a communication-efficient framework for distributed on-device LLM inference over wireless networks [27] partitions the weight tensors of an LLM across multiple devices and designs an over-the-air all-reduce scheme. Besides, Jupiter [28] introduces a fast and resource-efficient pipelined architecture, which exploits the different characteristics of the prefill and decoding phases for collaborative LLM inference on edge devices.

However, these works are fundamentally different from this paper. First, they mainly consider model partitioning, tensor partitioning, or inference assignment, and do not exploit the internal expert similarity or the depth redundancy of MoE models as a deployment and execution dimension. Second, they assume the exact model execution, i.e., when a model component is assigned to a remote node, the corresponding token representation must still be transmitted to that node for execution. Third, although these works improve the efficiency of distributed inference, they only accelerate the necessary transmissions, rather than considering whether it is necessary to execute a layer, or the remaining layers, at all. Therefore, they cannot achieve the additional communication reduction studied in this paper.

### *B. MoE Serving Optimization and Communication Reduction*

The efficiency of MoE inference has attracted increasing attention in recent years, because the sparsity of MoE models introduces not only computation challenges but also new communication and memory challenges. Specifically, when the activated experts of the tokens are distributed across different devices or servers, the expert dispatch and communication may become a dominant bottleneck.

Several works optimize MoE inference through expert offloading, caching, prefetching, replication, and communication scheduling. AdapMoE [29] adopts an adaptive sensitivity-based expert gating and management method to reduce the on-demand loading overhead of the sparsely activated experts. SlimCaching [30] investigates the edge caching of MoE experts for distributed low-latency inference, which stores the frequently activated experts on devices and caches the remaining experts across the edge network. Diff-MoE [31] proposes a priority-driven differential expert caching hierarchy, together with a lightweight predictor, to improve the batched MoE inference. PROBE [32] further co-balances the computation and communication through real-time predictive prefetching, dynamic expert replication, and token assignment. MegaScale-Infer [33] disaggregates the attention and expert modules and adopts the disaggregated expert parallelism for efficient large-scale MoE serving. Besides, some recent works also investigate the distributed or wireless MoE deployment at the edge: WDMoE [34] proposes a wireless distributed MoE architecture, which collaboratively deploys the experts across base-station edge servers and mobile devices, and MoE2 [35] optimizes the collaborative inference for edge LLMs under the energy and latency constraints. Together with the sparsely-activated MoE models themselves [4] and the recent MoE survey [5], these works demonstrate that communication-aware expert management is important to the efficiency of MoE inference.

However, the above works still differ from this paper in several key aspects. First, their objective is mainly to access the exact target experts more efficiently through better caching, replication, prefetching, dispatch, or scheduling. In contrast, this paper considers whether the remote target-expert invocation itself can be avoided. Second, most existing works assume that the routed target expert is the only valid execution and must always be executed, so they neither exploit the functionally similar local substitutes nor allow a low-importance layer to be skipped or the inference to terminate early. Therefore, they reduce the communication by transporting the tokens faster, whereas this paper reduces the communication by eliminating the layer executions and the cross-server transmissions they would otherwise trigger.

*C. Layer Skipping and Early Exit for Efficient Inference*

Another line of research reduces the inference cost along the depth dimension, which is motivated by the observation that the per-token computation of deep models is highly redundant. For the layer-level redundancy, ShortGPT [20] introduces a block-influence metric and shows that many layers of LLMs can be removed with negligible accuracy loss. Mixture-of-Depths [21] learns to dynamically allocate the computation to a subset of tokens at each layer, and bypasses the remaining tokens through the residual path under a fixed compute budget. For the early exit, Confident Adaptive Language Modeling [22] dynamically allocates different amounts of computation per token and exits the decoding once a calibrated confidence measure is met, with provable quality guarantees, and LayerSkip [36] combines the layer-dropout training with an early-exit loss and self-speculative decoding to enable accurate exit at earlier layers.

However, these works differ from this paper in two essential aspects. First, they are designed for single-device inference and aim to reduce the FLOPs or the decoding steps. They are not communication-aware and are not formulated for distributed multi-server edge MoE serving, so they ignore the fact that skipping or exiting a layer can additionally remove a cross-server transmission. Second, they treat layer skipping and early exit as standalone compute-saving mechanisms, and do not couple them with the server-expert selection along the token trajectory, nor enforce a per-user quality-degradation budget jointly with the server selection and the substitute-execution candidates. In contrast, this paper jointly exploits communication-aware layer skipping and confidence-based early exit, coordinated with substitute-aware server-expert selection, to remove both the unnecessary layer executions and their induced cross-server transmissions, while keeping the inference-quality degradation within a controllable budget.

## III. Network Model and Problem Statement

In this section, we first briefly introduce the distributed edge MoE deployment and the similarity-based substitute execution model established in our previous work [37]. Then, we introduce the expert importance and skip-layer model, the confidence-aware early-exit model, the inference delay model, and the unified quality degradation model that are used in this paper. Based on these models, the problem solved in this paper is described in detail.

*A. Network Model and Distributed MoE Deployment*

We consider the same distributed edge intelligence system as in [37]. Let $D = \{d_1, d_2, \dots, d_U\}$ denote the set of mobile users and $S = \{S_1, S_2, \dots, S_M\}$ denote the set of edge servers. For user $d_i$, its associated edge server is $S_{a(i)}$, which receives the request and serves as the initial execution server. The edge servers are geographically distributed and connected through ordinary Internet links. Let $B_{m,n}$ and $\tau_{m,n}$ denote the available bandwidth and the propagation-plus-queuing delay between $S_m$ and $S_n$, both of which differ with different server pairs. The servers are heterogeneous, where $C_m$ and $F_m$ denote the available GPU memory capacity and computation capability of $S_m$, respectively.

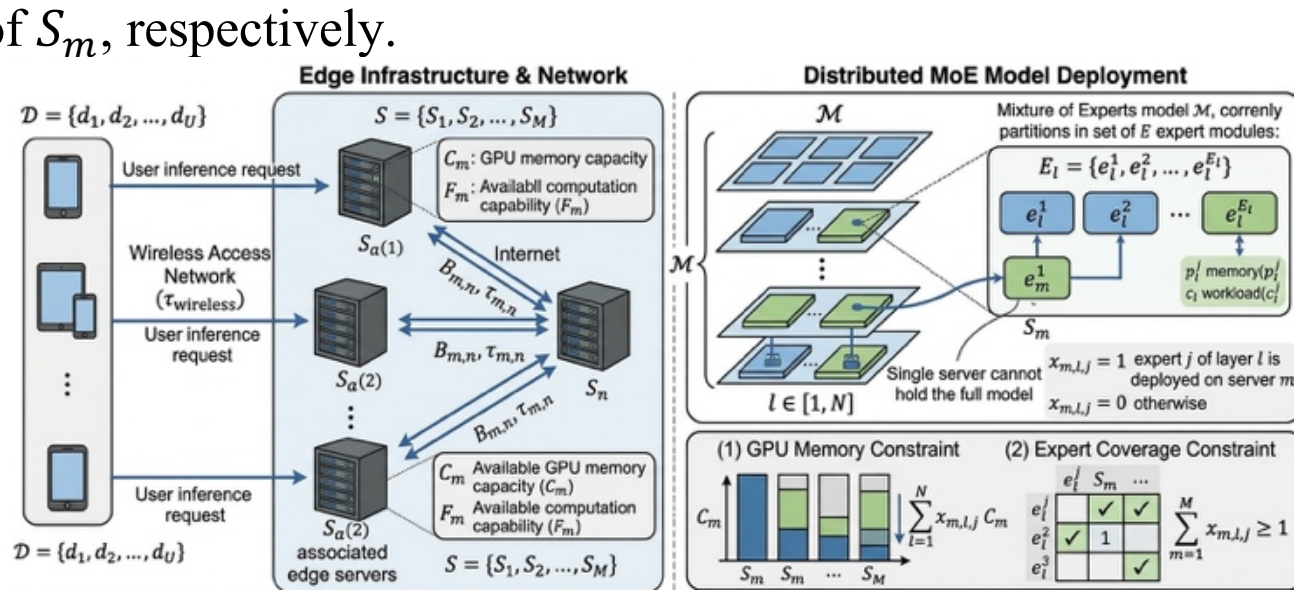

Fig.1 Network model

The MoE model deployed at the edge is denoted by $\mathcal{M}$, which contains $N$ MoE layers. In the $l$-th MoE layer, the expert set is $E_l = \{e_l^1, e_l^2, \dots, e_l^{E_l}\}$, where expert $e_l^j$ uses GPU memory $p_l^j$ and incurs computation workload $c_l^j$ for processing a token representation. Since the whole model cannot be deployed on a single server, the experts are distributed across the edge servers. Let $x_{m,l,j} = 1$ indicate that expert $e_l^j$ is deployed on $S_m$. The deployment satisfies the GPU memory constraint and the coverage constraint:

$$\sum_{l=1}^{N} \sum_{j=1}^{E_l} x_{m,l,j}\, p_l^j \leq C_m, \quad \forall S_m \in \mathcal{S} \tag{1}$$

$$\sum_{m=1}^{M} x_{m,l,j} \geq 1, \forall l \in [1, N], j \in [1, E_l] \tag{2}$$

To reduce the cross-server transmission caused by remote expert invocation, our previous work [37] groups functionally similar same-layer experts and disperses them across edge servers, so that each server can preserve local substitute candidates. For a target expert $e_l^r$ selected by the router, let $V_{l,r} = \{e_l^r\} \cup R_{l,r}$ denote its feasible execution expert set, where $R_{l,r}$ is the set of feasible substitute experts whose router-induced similarity with $e_l^r$ exceeds a layer-aware threshold [37]. When a substitute expert $e_l^k \in R_{l,r}$ is executed instead of the target expert, a substitution quality loss is introduced:

$$Q_l(r_l(q), k) = \begin{cases} 0, & k = r_l(q), \\ \phi\left(\frac{1 - \mathrm{Sim}_{\,l}(r_l(q), k)}{2}\right), & k \neq r_l(q), \end{cases} \tag{3}$$

where $\phi(\cdot)$ is a non-decreasing function. The detailed grouping, deployment, and similarity computation follow [37].

*B. Expert Importance and Skip-Layer Model*

For an inference request of user $d_i$, let $Q_i$ denote its token set, and $r_l(q)$ denote the target expert selected by the router for token $q$ in the $l$-th MoE layer. Since the routers are small and replicated on all servers [37], the routing decision of each layer is available at the server where the token currently resides, before the expert is actually executed. We exploit the router output not only to obtain the target expert, but also to estimate how important this layer is for the token.

At layer $l$, the router activates a top-k expert set for token $q$, denoted $E_{l(q)}$, where $g_l^k(q)$ is the gating weight of the k-th activated expert. Models with a shared expert treat the shared expert as always locally resident; only routed experts in $E_{l(q)}$ are subject to placement and selection. The layer importance is the aggregate gating mass of the activated set:

$$I_{i,q,l} = \sum_{k \in E_{l(q)}} g_l^k(q), I_{i,q,l} \in (0,1] \tag{4}$$

A larger $I_{i,q,l}$ means the router is more confident that this specific expert is decisive for the token, so omitting it would distort the representation more. Conversely, a small $I_{i,q,l}$ indicates that the activated expert contributes little, and the layer can be bypassed through the residual path with limited impact. Since $I_{i,q,l}$ is computed from the router output alone, the importance of a layer is known without executing its expert, which makes the skip decision feasible at runtime.

Let the skip decision be $\delta_{i,q,l} \in \{0,1\}$, where $\delta_{i,q,l} = 1$ means the expert computation of token $q$ in the $l$-th MoE layer is skipped and the token representation is forwarded through the identity (residual) path. When a layer is skipped, the predicted quality degradation is modeled as a non-decreasing function of its importance:

$$\Lambda_{i,q,l} = \rho\left(I_{i,q,l}\right) \tag{5}$$

where $\rho(\cdot)$ is a non-decreasing function. The skip-layer mechanism is meaningful only when it removes cross-server transmission. If a feasible expert in $V_{l,r_l(q)}$ is already available on the current server, the layer can be processed locally by exact or substitute execution at almost no transmission cost, and there is no need to skip it. Therefore, skipping is preferentially considered for a layer whose feasible experts all reside on remote servers, i.e., whose execution would otherwise require the token to be transmitted across servers. In this way, a skip simultaneously saves the computation of an unimportant expert and the transmission that would be needed to reach it.

*C. Confidence-Aware Early-Exit Model*

Besides skipping individual layers, a token may not need to traverse all $N$ layers. Once the representation of a token is already sufficient for the task requirement, the remaining layers can be terminated, which avoids both the residual computation and any cross-server transmission on the remaining trajectory.

To describe the depth that each token actually traverses, we introduce the active indicator $y_{i,q,l} \in \{0,1\}$, where $y_{i,q,l} = 1$ means token $q$ is still active when entering the $l$-th MoE layer. Since a terminated token is never reactivated, the active indicator is monotonically non-increasing along the depth:

$$y_{i,q,1} = 1, y_{i,q,l} \geq y_{i,q,l+1}, \forall l \in [1, N-1] \tag{6}$$

The exit layer of token $q$ is then $L_{i,q} = \sum_{l=1}^{N} y_{i,q,l}$, and layers $l > L_{i,q}$ are not executed.

Because the ground-truth output is not available during inference, the sufficiency of a token cannot be measured by accuracy directly. Instead, we attach a lightweight exit head $h_l(\cdot)$ to each layer. Let $\mathbf{h}_{i,q,l} \in \mathbb{R}^d$ denote the hidden representation of token $q$ of user $d_i$ at the output of the $l$-th MoE layer, where $d$ is the model hidden dimension. The exit head maps this representation to a confidence score:

$$\xi_{i,q,l} = h_l(\mathbf{h}_{i,q,l}) \in [0,1] \tag{7}$$

where $\xi_{i,q,l}$ reflects how stable and decisive the current representation is, and is consistent with established early-exit Transformer practice. Let $p_J^i$ denote the task confidence requirement of user $d_i$. Token $q$ is allowed to exit after the $l$-th MoE layer only if its confidence already meets the requirement:

$$y_{i,q,l+1} = 0 \Rightarrow \xi_{i,q,l} \geq p_J^i \tag{8}$$

Equation (8) guarantees that early exit is triggered only when the token is confident enough, so that early exit is quality-preserving by construction rather than a source of uncontrolled error.

It should be noted that the skip-layer mechanism and the early-exit mechanism may act on the same layer, and their interaction must be specified to keep the confidence judgment well defined. When the $l$-th MoE layer is skipped ($\delta_{i,q,l} = 1$), the expert is not actually executed, and the hidden representation $\mathbf{h}_{i,q,l}$ is merely propagated from the input through the identity (residual) path rather than refined by an expert. In this case the confidence score produced by the exit head does not reflect any newly performed computation and is therefore unreliable for exit judgment. To avoid making an exit decision on a representation that has not been updated, the exit judgment is performed only at layers that are actually executed, while a skipped layer never triggers early exit:

$$\delta_{i,q,l} = 1 \Rightarrow y_{i,q,l+1} = y_{i,q,l} \tag{9}$$

According to (9), when a layer is skipped, the active state is passed unchanged to the next layer, so that skipping and early exit cannot be triggered simultaneously at the same layer. This rule ensures that the confidence condition (8) is always evaluated on a representation refined by an executed expert, and that the two depth-reduction mechanisms remain mutually consistent along the token trajectory. When a token is skipped or exited, its representation is propagated to the dependent computations through the residual path, so that the dependencies of the subsequent tokens remain well defined. In this paper, we focus on the placement and invocation of MoE experts, which dominate the memory occupation and cross-server transmission overhead.

*Attention placement*. Attention sublayers, routers, and shared experts are dense and comparatively small; one copy of each is deployed on every participating server, with the KV cache co-located on the token's current server. Consequently, attention is always computed locally and never triggers cross-server transmission; all cross-server traffic is induced by routed-expert (FFN) execution when a required expert is unavailable locally. A layer skip bypasses only the FFN sublayer, while the attention sublayer of that block is still computed locally. Under this placement, the latency model treats local attention as a fixed per-layer computation term and attributes all cross-server delay to remote expert execution.

*D. Token Trajectory and Inference Delay Model*

Let $u_{i,q,l,m,k} = 1$ indicate that token $q$ of user $d_i$ is processed by expert $e_l^k$ on server $S_m$ in the $l$-th MoE layer. A server-expert pair is selected only for a layer that is active and not skipped. Let $\varepsilon_{i,q,l} = y_{i,q,l}\,(1 - \delta_{i,q,l})$ denote the executed indicator. Then,

$$\sum_{m=1}^{M} \sum_{k=1}^{E_l} u_{i,q,l,m,k} = \varepsilon_{i,q,l}, \forall d_i, q, l \tag{10}$$

$$u_{i,q,l,m,k} \leq x_{m,l,k}, u_{i,q,l,m,k} = 0 \ \text{ if } e_l^k \notin V_{l,r_l(q)} \tag{11}$$

The execution-server indicator is $\mu_{i,q,l,m} = \sum_{k=1}^{E_l} u_{i,q,l,m,k}$. Let $\ell_{i,q,l}$ denote the server on which the token resides before entering the $l$-th MoE layer. The initial location is the access edge server, and the location evolves according to whether the layer is executed or skipped:

$$\ell_{i,q,1} = a(i) \tag{12}$$

$$\ell_{i,q,l+1} = \begin{cases} \arg_m(\mu_{i,q,l,m} = 1), & \varepsilon_{i,q,l} = 1 \\ \ell_{i,q,l}, & \delta_{i,q,l} = 1 \end{cases} \tag{13}$$

According to (12), a skipped layer keeps the token at its current server, so it incurs no transmission. The transmission delay of token $q$ in the $l$-th MoE layer is only paid when the layer is executed on a server different from the current location:

$$T_{i,q,l}^{\text{trans}} = \sum_{m=1}^{M} \mu_{i,q,l,m} \mathbb{1}\left( m \neq \ell_{i,q,l} \right) \left( \frac{z_l}{B_{\ell_{i,q,l},m}} + \tau_{\ell_{i,q,l},m} \right) \quad (14)$$

where $z_l$ is the size of the intermediate token representation and $\mathbb{1}(\cdot)$ is the indicator function. The computation delay is incurred only by executed layers:

$$T_{i,q,l}^{\text{comp}} = \sum_{m=1}^{M} \sum_{k=1}^{E_l} u_{i,q,l,m,k} \cdot \frac{c_l^k}{F_m} \quad (15)$$

A skipped layer ($\delta_{i,q,l} = 1$) and an inactive layer ($y_{i,q,l} = 0$) contribute zero to both (14) and (15). Therefore, the total delay of token $q$ and the total inference delay of user $d_i$ are:

$$T_i = \sum_{q \in Q_i} \sum_{l=1}^{N} T_{i,q,l} \quad (16)$$

where $T_{i,q,l} = T_{i,q,l}^{\text{trans}} + T_{i,q,l}^{\text{comp}}$. Equation (16) shows the source of the additional reduction in this paper: a skipped remote layer removes one transmission term in (14), and early exit removes all the delay terms of the layers beyond $L_{i,q}$.

*E. Unified Quality Degradation Model*

In this paper, the inference quality of a token is affected by three approximation behaviors: substitute execution, layer skipping, and early exit. By the confidence condition (8), early exit is only triggered when the task requirement is already met, so it is modeled as a hard feasibility constraint rather than a quality loss. The quality degradation that must be budgeted therefore comes from substitution and skipping. For token $q$ in the $l$-th MoE layer, the per-layer degradation is:

$$D_{i,q,l} = \underbrace{\sum_{m=1}^{M} \sum_{k=1}^{E_l} u_{i,q,l,m,k} Q_l(r_l(q), k)}_{\text{substitution}} + \underbrace{\delta_{i,q,l} \Lambda_{i,q,l}}_{\text{skip}} \quad (17)$$

The first term is the substitution quality loss inherited from [37], and the second term is the skip-induced degradation defined in (5). Both degradation sources are expressed on a common probabilistic scale. $Q_l(r, k)$ is the fraction of calibration tokens whose final prediction changes when expert $e_l^r$ is replaced by substitute $e_l^k$; $\Lambda_l(l)$ is the fraction whose prediction changes when layer $l$ is skipped at importance $I$. Both lie in [0,1] on the same calibration distribution, so they share a common unit and can be summed within the budget $\overline{D}(i)$.

The total quality degradation of user $d_i$ is:

$$D_i = \sum_{q \in Q_i} \sum_{l=1}^{N} D_{i,q,l} \quad (18)$$

To guarantee inference quality, the total degradation of each user should not exceed an acceptable budget:

$$D_i \leq D_i^{\max}, \forall d_i \in D \quad (19)$$

The above model indicates that the server selection, expert selection, skip decision, and exit decision are jointly related to inference delay and quality degradation. For example, if the feasible experts of the next MoE layer all reside on remote servers, the token can be transmitted to a remote server for exact or substitute execution, which preserves quality but increases transmission delay; alternatively, if the layer is of low importance, the token can skip the layer and stay on the current server, which removes the transmission delay but consumes part of the quality budget. Moreover, once the token confidence reaches the task requirement, terminating the remaining layers removes both the residual computation and the remaining cross-server transmission. Therefore, it is necessary to jointly consider inference delay and quality degradation when determining the deployment, the token trajectory, the skip decision, and the exit decision.

*F. Problem Statement*

From the above subsections, we obtain the inference delay $T_i$ and the quality degradation $D_i$ of user $d_i$. The purpose of this paper is to minimize the inference delay and the quality degradation by jointly optimizing the expert deployment strategy $\boldsymbol{X}$, the server-expert selection strategy $\boldsymbol{U} = \{u_{i,q,l,m,k}\}$, the skip-layer strategy $\boldsymbol{\Delta} = \{\delta_{i,q,l}\}$, and the early-exit strategy $\boldsymbol{Y} = \{y_{i,q,l}\}$, where the expert grouping and the feasible substitute sets are obtained as in [37]. Since the inter-server links and the computation resources are shared, the per-server workload within a scheduling slot $\Delta t$ is bounded. Let $D(t)$ denote the set of requests served in slot $t$. The computation workload assigned to $S_m$ in slot $t$ is:

$$W_m(\boldsymbol{X}, \boldsymbol{U}; t) = \sum_{d_i \in D(t)} \sum_{q \in Q_i} \sum_{l=1}^{N} \sum_{k=1}^{E_l} u_{i,q,l,m,k} \cdot c_l^k \quad (20)$$

Then, the problem P0 is expressed as:

**P0**: $\min\{\sum_{i=1}^{U} T_i, \sum_{i=1}^{U} D_i\}$

s.t.:

$$\sum_{l=1}^{N} \sum_{j=1}^{E_l} x_{m,l,j}\, p_l^j \leq C_m, \forall S_m \in \mathcal{S} \quad \text{(c.1)}$$

$$\sum_{m=1}^{M} x_{m,l,j} \geq 1, \forall l \in [1, N], j \in [1, E_l] \quad \text{(c.2)}$$

$$x_{m,l,j} \in \{0,1\}, u_{i,q,l,m,k} \in \{0,1\} \quad \text{(c.3)}$$

$$\delta_{i,q,l} \in \{0,1\}, y_{i,q,l} \in \{0,1\} \quad \text{(c.4)}$$

$$y_{i,q,1} = 1, y_{i,q,l} \geq y_{i,q,l+1}, \forall l \in [1, N-1] \quad \text{(c.5)}$$

$$\delta_{i,q,l} = 1 \Rightarrow y_{i,q,l+1} = y_{i,q,l}, \forall d_i, q, l \quad \text{(c.6)}$$

$$\sum_{m=1}^{M} \sum_{k=1}^{E_l} u_{i,q,l,m,k} = y_{i,q,l}\,(1 - \delta_{i,q,l}), \forall d_i, q, l \quad \text{(c.7)}$$

$$u_{i,q,l,m,k} \leq x_{m,l,k}, u_{i,q,l,m,k} = 0 \text{ if } e_l^k \notin V_{l,r_l(q)} \quad \text{(c.8)}$$

$$y_{i,q,l+1} = 0 \Rightarrow \xi_{i,q,l} \geq p_J^i, \forall d_i, q, l \quad \text{(c.9)}$$

$$D_i \leq D_i^{\max}, \forall d_i \in D \quad \text{(c.10)}$$

$$W_m(\boldsymbol{X}, \boldsymbol{U}; t) \leq F_m \Delta t, \forall S_m \in S, \forall t \quad \text{(c.11)}$$

In P0, constraints (c.1) to (c.3) are the deployment and binary-selection constraints inherited from [37]. Constraint (c.4) means that the skip and exit decisions are binary. Constraint (c.5) guarantees that the active indicator is monotonically non-increasing along the depth, so that a terminated token is never reactivated. Constraint (c.6) is the skip-exit coordination constraint: when a layer is skipped, its representation is not refined by an executed expert, so the active state is passed unchanged to the next layer and early exit cannot be triggered at the same layer. Constraint (c.7) means that a server-expert pair is selected if and only if the layer is active and not skipped. Constraint (c.8) guarantees that an executed expert is deployed on the selected server and is either the target expert or a feasible substitute expert. Constraint (c.9) guarantees that early exit is triggered only when the task confidence requirement is satisfied. Constraint (c.10) bounds the total quality degradation of each user, and (c.11) bounds the per-server workload within each scheduling slot.

This problem is difficult to solve for the following reasons. First, the objectives are conflicting: skipping a remote layer or exiting early reduces transmission and computation delay, but skipping consumes the quality budget, while exact remote execution preserves quality at the cost of extra transmission delay. Second, the deployment, selection, skip, and exit decisions are all discrete, which makes the solution space extremely large. Third, the decisions of different layers are coupled along the token trajectory: the execution server selected in one layer determines the source server of the next executed layer, and skipping or exiting at one layer changes both the residing server and the remaining trajectory. Fourth, the importance $I_{i,q,l}$ and the confidence $\xi_{i,q,l}$ of a layer are only revealed when the token reaches that layer, so the skip and exit decisions must be made in an online and predictive manner. Therefore, finding the optimal solution of P0 is difficult.

Since the objectives in P0 are different, a weight-based approach is introduced. Let $\widetilde{T}_i = T_i / T_i^{\text{ref}}$ and $\widetilde{D}_i = D_i / D_i^{\max}$ denote the normalized inference delay and the normalized quality degradation of user $d_i$, where $T_i^{\text{ref}}$ is the inference delay under the pure full-depth exact-execution strategy on the same deployment. The utility function is:

$$U(\boldsymbol{X}, \boldsymbol{U}, \boldsymbol{\Delta}, \boldsymbol{Y}) = \omega_T \sum_{i=1}^{U} \widetilde{T}_i + \omega_D \sum_{i=1}^{U} \widetilde{D}_i \tag{21}$$

where $\omega_T$ and $\omega_Q$ are the weights of inference delay and substitution quality loss, respectively, and $\omega_T + \omega_Q = 1$.

Thus, P0 is transformed into the single-objective problem:

$$\textbf{P1}: \quad \min_{\boldsymbol{X}, \boldsymbol{U}, \boldsymbol{\Delta}, \boldsymbol{Y}} U(\boldsymbol{X}, \boldsymbol{U}, \boldsymbol{\Delta}, \boldsymbol{Y})$$

According to (21), a larger $\omega_T$ favors low latency through more aggressive skipping and earlier exit, while a larger $\omega_D$ favors quality preservation through more exact and full-depth execution. Therefore, the weight-based approach can flexibly achieve the tradeoff between inference delay and inference quality.

## IV. Offline Exit-Head Training and Communication Aware Deployment

In Section III, the joint optimization problem P1 has been formulated. In this section, we therefore present the offline procedure that produces these calibration quantities and refines the deployment strategy $X$ in a communication-aware manner.

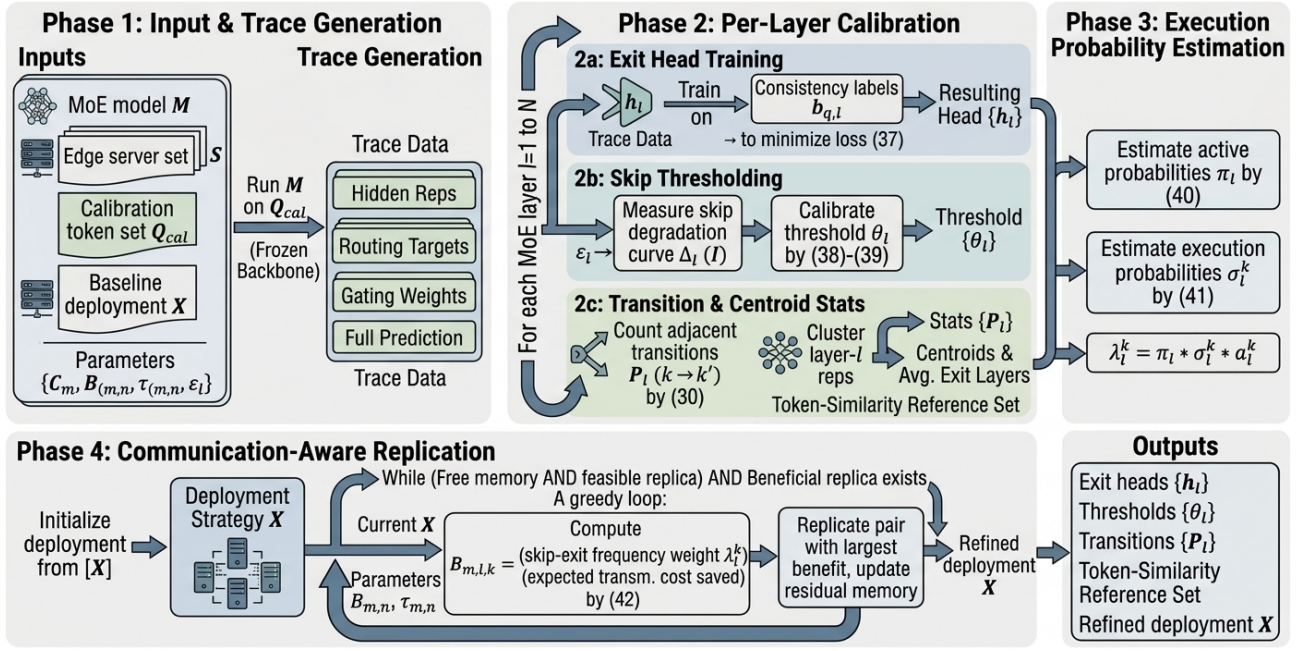

Fig.2. Offline exit-head training and communication aware deployment

### A. Exit-Head Training

The early-exit decision relies on the confidence score $\xi_{i,q,l} = h_l(\mathbf{h}_{i,q,l})$ defined in (7). For the exit decision to be quality-preserving as required by constraint (c.9), the confidence score must be a faithful indicator of whether the current representation already yields the same prediction as the full-depth model. We therefore train one lightweight exit head per layer on the calibration set.

The exit head $h_l(\cdot)$ is a single linear layer followed by a sigmoid activation, which maps the hidden representation $\mathbf{h}_{i,q,l} \in \mathbb{R}^d$ to a scalar in $[0,1]$. Its parameter count is negligible compared with an expert, so the head incurs almost no memory and computation overhead. For each calibration token $q \in \mathcal{Q}_{\text{cal}}$, let $\hat{o}_{q,l}$ denote the prediction obtained by applying the model output head to $\mathbf{h}_{q,l}$, and let $o_q^*$ denote the full-depth prediction of the same token. The supervision label indicates whether the layer-$l$ representation is already consistent with the full-depth output:

$$b_{q,l} = \mathbb{1}(\hat{o}_{q,l} = o_q^*) \tag{22}$$

Each exit head is then trained to predict this consistency by minimizing the binary cross-entropy loss over the calibration set:

$$\mathcal{L}_l = -\frac{1}{|\mathcal{Q}_{\text{cal}}|} \sum_{q \in \mathcal{Q}_{\text{cal}}} \left[ b_{q,l} \log \xi_{q,l} + (1 - b_{q,l}) \log (1 - \xi_{q,l}) \right] \tag{23}$$

The backbone model is frozen during this stage, and only the exit heads are trained, so the original routing behavior and expert outputs are not altered. Consequently, the confidence score learns to approximate the probability that exiting at layer $l$ preserves the full-depth prediction. This is exactly the semantic required by constraint (c.9): when $\xi_{i,q,l} \geq p_J^i$, the representation is consistent with the full-depth output with probability at least $p_J^i$, so the exit is quality-preserving by construction rather than a source of uncontrolled error.

Note that the proposed exit-head training does not fine-tune the backbone model. The MoE experts, routers, and attention layers are all frozen throughout, and only the per-layer linear exit heads are trained on cached hidden representations. As a result, no gradient is back-propagated through the backbone, no optimizer state of the backbone is maintained, and the original routing decisions and expert outputs remain bit-identical to the pre-trained model. The total number of trainable parameters introduced by all exit heads is approximately $N \cdot d$, which is negligible compared with a single expert. Consequently, both the offline training cost and the runtime memory and latency overhead introduced by the exit heads are marginal.

### B. Layer Importance Threshold Calibration

The skip decision $\delta_{i,q,l}$ admits the skip action only when the layer importance in (4) is below a layer-specific threshold $\theta_l$. The role of $\theta_l$ is to bound the degradation introduced by skipping, so that the runtime budget enforcement in Algorithm 2 (in the next Section) is rarely the binding constraint and the per-step skip cost stays meaningful. We calibrate $\theta_l$ offline from the empirical relationship between layer importance and skip-induced degradation.

On the calibration set, for each layer $l$ we measure the actual quality degradation caused by skipping the layer for tokens with different importance values. Let $\Delta_l(I)$ denote the average degradation observed when layers with importance in a small neighborhood of $I$ are skipped, where the degradation is measured as the consistency drop between the skipped-path prediction and the full-depth prediction and $\Lambda_{i,q,l} = \Delta_l(I_{i,q,l})$. Since a higher importance corresponds to a larger degradation, $\Delta_l(I)$ is also non-decreasing in $I$, which is consistent with the modeling assumption of (5). The threshold $\theta_l$ is then selected as the largest importance value whose skip degradation does not exceed a per-layer tolerance $\varepsilon_l$:

$$\theta_l = \max \{ I \in [0,1] \mid \Delta_l(I) \leq \varepsilon_l \} \tag{24}$$

The per-layer tolerance is derived from the user budget so that the offline threshold and the online budget are mutually consistent. Letting $\bar{D}$ denote the average per-token budget over all users and $\bar{\kappa}$ the expected number of skipped layers per token observed on the calibration set, the tolerance is set as:

$$\varepsilon_l = \frac{\bar{D}}{\bar{\kappa}} \tag{25}$$

Equation (25) distributes the per-token budget evenly across the expected skipped layers, so that a token that skips $\bar{\kappa}$ layers,

each within tolerance $\varepsilon_l$, accumulates an expected degradation no larger than $\bar{D}$. This makes the offline threshold a conservative pre-filter: it removes high-importance layers from the skip candidates before the runtime budget check in (19), so that the two mechanisms work in the same direction and never contradict each other. We emphasize that $\theta_l$ is only an admission pre-filter; the hard guarantee of constraint (c.10) is still enforced online by the budget condition in (19), so the quality constraint holds regardless of any calibration mismatch.

*C. Communication-Aware Redundant Deployment*

In our previous work [37], it guarantees that every expert is deployed at least once, so the coverage constraint (c.2) and the memory constraint (c.1) are satisfied; when edge servers still have free memory after the baseline deployment, important experts can be replicated to further reduce runtime cross-server transmission. However, in this paper the skip and exit mechanisms change how often each expert is actually executed, so replicating an expert that is frequently skipped or that lies beyond the typical exit layer wastes memory without reducing transmission. We therefore redefine the redundancy benefit to be aware of the skip and exit behavior.

Let $\pi_l$ denote the probability that a token is still active when entering layer $l$, estimated from the calibration exit distribution:

$$\pi_l = \frac{1}{|\mathcal{Q}_{\text{cal}}|}\sum_{q\in\mathcal{Q}_{\text{cal}}} \mathbb{1}\left(L_q \geq l\right) \tag{26}$$

Since the active indicator is monotonically non-increasing along the depth by (6), $\pi_l$ is non-increasing in $l$, which means deeper experts are reached by fewer tokens. Let $\sigma_l^k$ denote the probability that expert $e_l^k$ is actually executed rather than skipped, given that the token is active at layer $l$. This is estimated as the fraction of calibration tokens routed to $e_l^k$ whose layer importance exceeds the skip threshold:

$$\sigma_l^k = \Pr\left(I_{q,l} > \theta_l \mid r_l(q) = k\right) \tag{27}$$

The expected execution frequency of expert $e_l^k$ is then $\lambda_l^k = \pi_l \cdot \sigma_l^k \cdot a_l^k$, where $a_l^k$ is the activation frequency of $e_l^k$ from [37]. This product form assumes the early-exit attrition $\pi_l$, skip attrition $\sigma_l^k$, and activation frequency $a_l^k$ are approximately independent. In practice they are weakly positively correlated, which slightly overestimates $\lambda_l^k$ for deep experts. This only inflates the redundancy benefit marginally and never produces an infeasible replication decision; the memory constraint is enforced independently.

Replicating $e_l^k$ on a server $S_m$ that currently must fetch the expert from a remote server saves the corresponding transmission for every token that resides on $S_m$ and executes $e_l^k$. Accordingly, the redundancy benefit of placing a replica of $e_l^k$ on $S_m$ is defined as:

$$B_{m,l,k} = \lambda_l^k \cdot \rho_{m,l} \cdot \left(\frac{z_l}{\bar{B}_m} + \bar{\tau}_m\right) \tag{28}$$

where $\rho_{m,l}$ is the fraction of tokens expected to reside on $S_m$ before layer $l$ (estimated from the access distribution and the baseline trajectories), and $\bar{B}_m$ and $\bar{\tau}_m$ are the average bandwidth and latency from $S_m$ to the servers currently hosting $e_l^k$. The benefit is large only when the expert is both frequently executed ($\lambda_l^k$ high, i.e., shallow and frequently activated and rarely skipped) and currently expensive to reach from $S_m$. Experts that are usually skipped (small $\sigma_l^k$) or lie beyond the typical exit depth (small $\pi_l$) receive a low benefit and are therefore not replicated, so the free memory is concentrated on the experts that genuinely cut transmission.

The replicas are then placed greedily: among all feasible (expert, server) pairs that satisfy the memory constraint (c.1), the pair with the largest redundancy benefit is replicated first, the residual memory is updated, and the process repeats until no feasible replica remains. Since replication only adds copies of already-deployed experts, the coverage constraint (c.2) and the memory constraint (c.1) continue to hold throughout, so the refined deployment $X$ remains feasible.

*F. Offline Algorithm*

Based on the above models, the offline exit-head training and communication-aware deployment algorithm is summarized in Table I.

Table I

| Alaogirtim1: Offline exit head training and communication aware deployment algorithm |
|---|
| **Input**: MoE model $\mathcal{M}$, edge server set $S$, calibration token set $\mathcal{Q}_{\text{cal}}$, baseline grouping $\mathcal{G}$ and deployment from [37], memory capacities $\{C_m\}$, bandwidths $\{B_{m,n}\}$, latencies $\{\tau_{m,n}\}$, per-layer skip tolerances $\{\varepsilon_l\}$. |
| **Output**: Exit heads $\{h_l\}$, importance thresholds $\{\theta_l\}$, transition statistics $\{P_l\}$, token-similarity reference set, refined deployment $X$. |
| 1. Run the full-depth model on $\mathcal{Q}_{\text{cal}}$ and record hidden representations, routing targets, gating weights, and full-depth predictions |
| 2. **for** each MoE layer $l = 1$ to $N$ **do** |
| 3. Build consistency labels $b_{q,l}$ by (22) |
| 4. Train exit head $h_l$ by minimizing the loss (23) with the backbone frozen |
| 5. Measure skip degradation curve $\Delta_l(I)$ and calibrate threshold $\theta_l$ by (24) to (25) |
| 6. Cluster layer-$l$ representations into centroids with average exit layers |
| 7. **end for** |
| 8. Estimate active probabilities $\pi_l$ by (26) and execution probabilities $\sigma_l^k$ by (27) |
| 9. Initialize $X$ from the baseline deployment of [37] |
| 10. **while** some server has free memory and a feasible replica exists **do** |
| 11. Compute redundancy benefit $B_{m,l,k}$ by (28) for all feasible (expert, server) pairs |
| 12. Replicate the pair with the largest benefit and update residual memory |
| 13. **end while** |
| 14. **return** $\{h_l\}$, $\{\theta_l\}$, $\{P_l\}$, token-similarity reference set, and $X$. |

The algorithm contains three parts.

**Lines 1-9: Per-layer calibration.** A single forward pass over the calibration set $\mathcal{Q}_{\text{cal}}$ is performed with the backbone frozen, and all intermediate quantities required by the offline procedure are recorded simultaneously, avoiding multiple passes over the data. For each MoE layer $l$, the hidden representations at the layer output, the routing targets, the gating weights, and the full-depth predictions are extracted and stored. The consistency label $b_{q,l}$ of each calibration token is then built by comparing the intermediate prediction obtained from $\mathbf{h}_{q,l}$ against the full-depth prediction, and the exit head $h_l$ is trained from scratch on these labels with only the head parameters updated. Because the backbone is strictly frozen throughout this stage, the routing logits and expert outputs are identical to those seen at runtime, so the transition counts and the calibration representations are not contaminated by any training-induced drift. For each layer, the skip degradation curve $\Delta_l(I)$ is measured by evaluating the consistency drop when tokens of different importance levels are forced to skip the layer, and the importance threshold $\theta_l$ is derived from this curve using the budget-derived tolerance $\varepsilon_l$ in (25), ensuring that the offline threshold and the online budget constraint point in the same direction.

**Lines 10-11: Execution-frequency estimation.** Once the per-layer calibration is complete, the effective execution frequency of each expert is estimated by composing two independently calibrated quantities. The active probability $\pi_l$ captures the depth-wise attrition of tokens due to early exit: it is the fraction of calibration tokens that are still active when entering layer $l$, and it is non-increasing in $l$ by construction, so deeper experts automatically receive a lower weight in the redundancy benefit. The execution probability $\sigma_l^k$ captures the additional attrition due to skip decisions: it is the fraction of calibration tokens routed to expert $e_l^k$ whose importance exceeds the calibrated threshold $\theta_l$, meaning that the expert would actually be executed rather than bypassed. The product $\lambda_l^k$ then gives the expected fraction of inference tokens that will actually invoke expert $e_l^k$ at runtime, accounting simultaneously for early exit and layer skipping. This quantity replaces the raw activation frequency used in [37], so that the subsequent replication step is not misled by an inflated frequency estimate for experts that are rarely reached or frequently skipped.

**Lines 12-16: Communication-aware replication.** Starting from the feasible baseline deployment inherited from [37], which already satisfies the coverage constraint (c.2) and the memory constraint (c.1), each edge server's residual memory is computed. The redundancy benefit $B_{m,l,k}$ in (28) is then evaluated for every (expert, server) pair that is not yet deployed and does not violate the memory constraint. The benefit explicitly weights the skip-exit-aware execution frequency $\lambda_l^k$ against the expected transmission cost saved by placing a replica on $S_m$, so that experts that are deeply buried beyond the typical exit layer, or that are almost always skipped, contribute negligible benefit regardless of how large their raw activation frequency is. The greedy loop selects the pair with the largest benefit, places the replica, and updates the residual memory of the corresponding server before the next iteration. The loop terminates when no feasible pair remains, either because no server has sufficient residual memory or because all beneficial replicas have already been placed. Because each iteration only adds a copy of an already-deployed expert and the memory check is enforced before placement, the constraints (c.1) to (c.3) remain satisfied at every iteration, and the final deployment $X$ is guaranteed feasible. The refined deployment and all calibration outputs are then returned together, forming the complete offline input to Algorithm 2.

## V. Communication Aware Online Skip Exit Server-Expert Selection

In Section IV, the offline expert grouping result $\mathcal{G}$ and the deployment strategy $X$ have been obtained. Based on these results, the remaining runtime tasks are threefold: determining which layers to skip, determining when to exit early, and selecting the server-expert pair for each executed layer. Therefore, in this section, we propose a communication-aware online strategy that jointly determines the skip decision, the exit decision, and the server-expert pair for each token at each layer, as illustrated in Fig. 3.

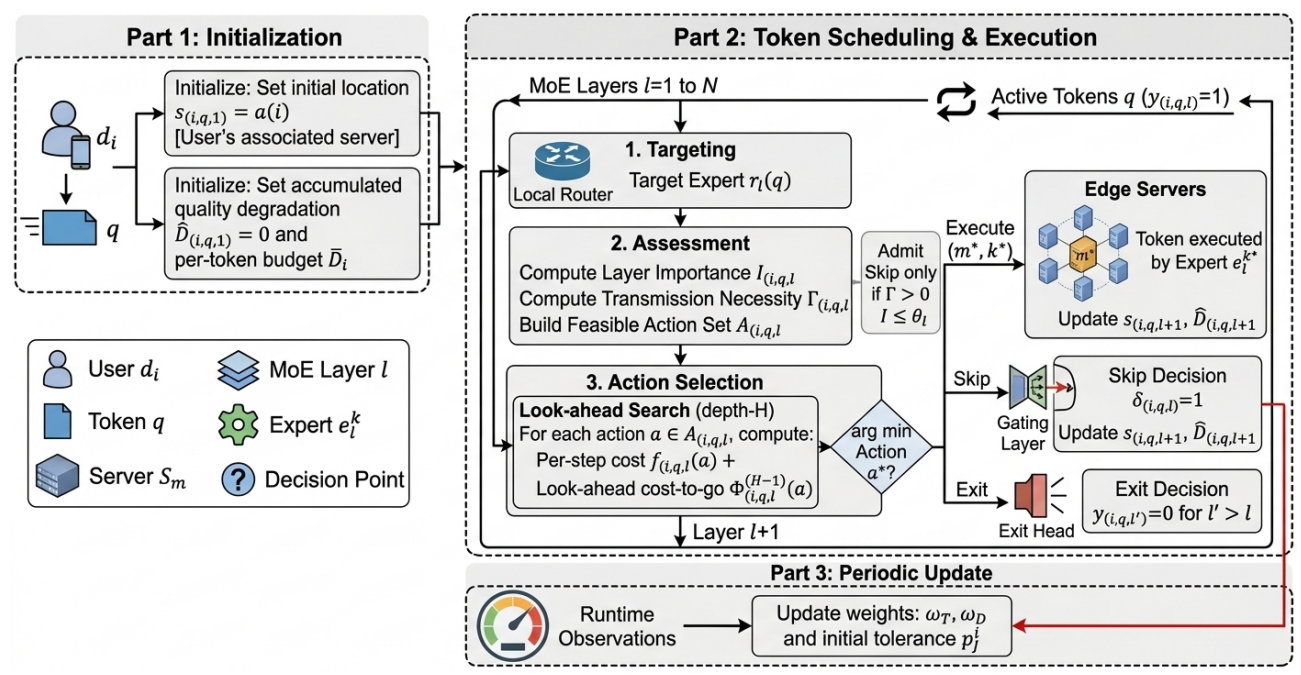

Fig.3 Communication aware online skip exit and server-expert selection

### *A. Online Decision State and Per-Step Cost*

The online strategy processes tokens layer by layer. For token $q$ of user $d_i$, let $s_{i,q,l}$ denote the server on which the token representation resides before entering the $l$-th MoE layer. The initial location is the access edge server of the corresponding user, i.e., $s_{i,q,1} = a(i)$, and the location is updated only when the layer is executed:

$$s_{i,q,l+1} = \begin{cases} m, & \text{if layer } l \text{ executed on } S_m \\ s_{i,q,l}, & \text{if layer } l \text{ skipped} \end{cases} \tag{29}$$

At the $l$-th MoE layer, three mutually exclusive actions are available for an active token ($y_{i,q,l} = 1$):

1) **Execute** on server $S_m$ by expert $e_l^k$ : the token is transmitted to $S_m$ if $S_m \neq s_{i,q,l}$, then processed by $e_l^k$.

2) **Skip**: the expert is bypassed through the residual path; the token stays on $s_{i,q,l}$ and no expert is invoked.

3) **Exit**: all layers from $l$ onward are terminated; this action is available only if $\xi_{i,q,l-1} \geq p_j^i$ (i.e., the confidence requirement is already satisfied after the previous executed layer).

When the token is executed on server $S_m$ by expert $e_l^k$, the per-step delay is the sum of the transmission delay from the current location to $S_m$ and the computation delay on $S_m$:

$$d_{i,q,l}(m,k) = \mathbb{1}(s_{i,q,l} \neq m)\left(\frac{z_l}{B_{s_{i,q,l},m}} + \tau_{s_{i,q,l},m}\right) + \frac{c_l^k}{F_m} \tag{30}$$

When the layer is skipped, no transmission is incurred and no computation is performed, so the skip delay is zero. The per-step cost of executing the server-expert pair $(m, k)$ is defined as the normalized weighted sum of delay and quality degradation:

$$f_{i,q,l}(m,k) = \omega_T \cdot \frac{d_{i,q,l}(m,k)}{d_i^{\text{ref}}} + \omega_D \cdot \frac{Q_l(r_l(q),k)}{\bar{D}_i} \tag{31}$$

where $d_i^{\text{ref}}$ is the per-step reference delay of user $d_i$ under full-depth local exact execution on the same deployment, and $\bar{D}_i$ is the per-token quality degradation budget defined in the next subsection. Both terms are dimensionless and share the same weights as the global utility (21), which ensures that the online per-step cost is consistent with the offline optimization objective.

The per-step cost of a skip action is defined analogously. Since a skip incurs zero delay but consumes a skip-induced degradation $\Lambda_{i,q,l} = \rho(I_{i,q,l}) = \Delta_l(I_{i,q,l})$, the skip cost is:

$$f_{i,q,l}^{\text{skip}} = \omega_D \cdot \frac{\Lambda_{i,q,l}}{\bar{D}_i} \tag{32}$$

An exit action incurs neither delay nor degradation beyond the current layer, so its incremental cost is zero. The exit action is therefore taken only when it is feasible (the confidence requirement is met) and the look-ahead cost of the remaining layers is positive, making exit strictly beneficial.

*B. Quality-Budget-Aware Feasible Action Set*

To enforce the per-user quality degradation constraint (c.10) in an online manner, the user-level budget $D_i^{\max}$ is evenly allocated across the tokens of user $d_i$:

$$\bar{D}_i = \frac{D_i^{\max}}{|Q_i|} \tag{33}$$

Let $\widehat{D}_{i,q,l}$ denote the accumulated quality degradation of token $q$ before the $l$-th MoE layer, initialized as zero. The runtime feasible action set of token $q$ at the $l$-th MoE layer is:

$$\mathcal{A}_{i,q,l} = \{(m,k) \mid x_{m,l,k} = 1, e_l^k \in V_{l,r_l(q)}, \widehat{D}_{i,q,l} + Q_l(r_l(q),k) \le \bar{D}_i\} \cup \{\text{skip} \mid \widehat{D}_{i,q,l} + \Lambda_{i,q,l} \le \bar{D}_i\} \cup \{\text{exit} \mid \xi_{i,q,l-1} \ge p_J^i\} \tag{34}$$

The first subset contains all execute candidates: expert $e_l^k$ must be deployed on $S_m$, must be the target expert or a feasible substitute, and the resulting accumulated degradation must not exceed the per-token budget. The second subset contains the skip action, which is feasible only if the skip-induced degradation does not exhaust the remaining budget. The third subset contains the exit action, which is feasible only if the confidence requirement has already been satisfied at the previous executed layer.

Since each expert is deployed on at least one server by constraint (c.2), the target expert $e_l^{r_l(q)}$ is always available somewhere, and its substitution quality loss is zero by definition. Therefore, the exact-execution candidate always satisfies the budget condition, which guarantees that $\mathcal{A}_{i,q,l}$ always contains at least one execute candidate and is never empty. The exit action is a hard feasibility gate controlled by the confidence score, not a soft preference, so it never forces premature termination. This construction ensures that the online feasible set is always consistent with the constraints (c.6) to (c.9) of P1.

*C. Importance-Guided Skip Decision with Look-Ahead*

The skip decision trades a fraction of the quality budget for the elimination of a cross-server transmission. To make this tradeoff principled rather than greedy, the strategy estimates the transmission cost that would be incurred if the layer were executed, and compares it against the skip cost.

**Transmission necessity score.** For the $l$-th MoE layer of token $q$, define the transmission necessity score as the minimum transmission delay over all execute candidates:

$$\Gamma_{i,q,l} = \min_{(m,k)\in\mathcal{A}_{i,q,l}^{\text{exec}}} \mathbb{1}(s_{i,q,l} \ne m)\left(\frac{z_l}{B_{s_{i,q,l},m}} + \tau_{s_{i,q,l},m}\right) \tag{35}$$

where $\mathcal{A}_{i,q,l}^{\text{exec}}$ is the execute subset of $\mathcal{A}_{i,q,l}$. If the minimum is zero, there exists a locally available execute candidate and the token need not be transmitted; the skip action in this case provides no transmission benefit and is thus suppressed. This formalizes the design principle in Section III-B that skipping is preferentially applied only to layers whose execution would require cross-server transmission.

**Importance threshold.** Even when a layer requires cross-server transmission, skipping is appropriate only if the layer importance is sufficiently low. The skip action is admitted into the decision only when $I_{i,q,l} \le \theta_l$, where $\theta_l$ is a layer-specific importance threshold calibrated offline (detailed in Section IV). When $I_{i,q,l} \le \theta_l$ is violated, the layer is deemed too important to skip, and the decision reduces to a server-expert selection among the execute candidates in $\mathcal{A}_{i,q,l}^{\text{exec}}$.

**Look-ahead skip evaluation.** Even when the skip action is admitted, its impact on future layers must be assessed. Skipping keeps the token on the current server $s_{i,q,l}$, which may be favorable for future layers if the subsequent experts are also hosted locally, or unfavorable if the skip merely defers a cross-server transmission to a later, more expensive hop. To capture this, the skip cost-to-go is estimated by the same look-ahead mechanism described in the next subsection, with the token location fixed at $s_{i,q,l}$ after the skip.

*D. Confidence-Guided Exit Decision with Token-Similarity Prediction*

The exit decision eliminates all remaining layers once the confidence of the current representation meets the task requirement. However, the confidence score $\xi_{i,q,l}$ is computed from the exit head $h_l(\cdot)$ and is only available after an executed layer; it is not available after a skipped layer by constraint (c.6). The exit decision therefore interleaves only with executed layers.

**Confidence-based exit gate.** After executing the $l$-th MoE layer, if $\xi_{i,q,l} \ge p_J^i$, the exit action becomes feasible. The decision of whether to exit or continue is then made by comparing the incremental cost of the remaining layers against the zero incremental cost of exit. If the estimated cost-to-go of the remaining layers is positive, exit is taken; if it is zero or negative (which occurs only when all remaining layers are locally available and nearly cost-free), execution continues.

**Token-similarity-based exit-point prediction.** To estimate the confidence of future layers without actually computing them, we exploit the observation that tokens with similar hidden representations at a given layer tend to exit at similar layers. Let $\mathcal{C}_l(q)$ denote the set of calibration tokens whose hidden representation at layer $l$ is closest to $\mathbf{h}_{i,q,l}$ under cosine similarity. The predicted exit layer of token $q$ given that it is still active at layer $l$ is:

$$\hat{L}_{i,q,l} = \text{round}\left(\frac{1}{|\mathcal{C}_l(q)|}\sum_{q'\in\mathcal{C}_l(q)} L_{q'}\right) \tag{36}$$

where $L_{q'}$ is the actual exit layer of calibration token $q'$. The predicted exit layer $\hat{L}_{i,q,l}$ is used to estimate the number of remaining executed layers and thus the expected future delay. If $\hat{L}_{i,q,l} \le l$, the prediction indicates that the token should have already exited, which strengthens the evidence for triggering exit at the current layer.

**Adaptive exit threshold.** In practice, the calibration-based prediction may be biased under distribution shift. To maintain calibration, the exit confidence threshold is adapted using a running estimate of the observed false-exit rate within the latest observation window:

$$p_J^i \leftarrow \Pi_{[\underline{p},1]}\left(p_J^i + \eta\left(\bar{\xi}_{\text{exit}} - p_J^i\right)\right) \tag{37}$$

where $\bar{\xi}_{\text{exit}}$ is the average confidence score at which exits were triggered in the observation window, $\eta$ is a small step size, and $\Pi_{[\underline{p},1]}(\cdot)$ projects the value onto $[\underline{p},1]$ with a minimum confidence floor $\underline{p}$ to prevent premature exit. The update pushes $p_J^i$ toward the observed exit confidence, so that the threshold self-corrects when the actual exit behavior deviates from the calibration.

*E. Look-Ahead Joint Selection*

For each active and non-exited token, the per-layer decision is made by selecting the action in $\mathcal{A}_{i,q,l}$ that minimizes the sum of the current-step cost and an estimated look-ahead cost-to-go over the following $H$ layers.

**Expert transition statistics.** Since the routing targets of future layers are not known before their representations are computed, the look-ahead relies on expert transition statistics collected on the calibration set. Let $P_l(k \to k')$ denote the empirical probability that a token routed to expert $e_l^k$ in the $l$-th MoE layer is subsequently routed to expert $e_{l+1}^{k'}$ in the next layer:

$$P_l(k \to k') = \frac{n_l(k \to k')}{\sum_{k''} n_l(k \to k'')} \quad (38)$$

where $n_l(k \to k')$ is the transition count observed on the calibration set. The transition statistics capture the layer-wise routing correlation, so that the look-ahead can estimate the distribution of future execution targets without computing future representations.

**Cost-to-go recursion.** The look-ahead cost-to-go of token $q$ being at server $S_m$ after taking action $(m,k)$ at the $l$-th MoE layer is defined by a depth-$H$ recursion. The base case is $\Phi_{i,q,l}^{(0)}(m,k) = 0$, and for $h \geq 1$:

$$\Phi_{i,q,l}^{(h)}(m,k) = \sum_{k'} P_l(k \to k') \cdot \min_{a' \in \mathcal{A}_{i,q,l+1}} \left[ f_{i,q,l+1}(a') \middle| s_{i,q,l+1} = m + \Phi_{i,q,l+1}^{(h-1)}(m,a') \right] \quad (39)$$

where $a'$ ranges over all feasible actions (execute, skip, or exit) in $\mathcal{A}_{i,q,l+1}$, and the per-step cost $f_{i,q,l+1}(a')$ is evaluated with the source location fixed at the server $S_m$ inherited from the current step. For the skip action $a' = \text{skip}$, the location remains $m$ and the cost is $f_{i,q,l+1}^{\text{skip}}$. For the exit action $a' = \text{exit}$, both the cost and the subsequent cost-to-go are zero. This recursive definition ensures that the look-ahead is always self-consistent with the token trajectory: the estimated future transmission is always measured from the server actually chosen at the current step, and the skip-exit coordination constraint (c.6) is respected within the recursion by suppressing exit at layers that are predicted to be skipped.

**Action selection.** Combining the current-step cost and the look-ahead cost-to-go, the action for token $q$ at the $l$-th MoE layer is selected as:

$$a^* = \arg\min_{a \in \mathcal{A}_{i,q,l}} \left[ f_{i,q,l}(a) + \Phi_{i,q,l}^{(H-1)}(a) \right] \quad (40)$$

When $a^*$ is an execute action $(m^*, k^*)$, the decision variable $u_{i,q,l,m^*,k^*} = 1$ is set, the token is executed on $S_{m^*}$ by expert $e_l^{k^*}$, and the token state is updated as:

$$s_{i,q,l+1} = m^*, \widehat{D}_{i,q,l+1} = \widehat{D}_{i,q,l} + Q_l(r_l(q), k^*) \quad (41)$$

When $a^*$ is skip, $\delta_{i,q,l} = 1$ is set, the token stays on $s_{i,q,l}$, and the accumulated degradation is updated as:

$$s_{i,q,l+1} = s_{i,q,l}, \widehat{D}_{i,q,l+1} = \widehat{D}_{i,q,l} + \Lambda_{i,q,l} \quad (42)$$

When $a^*$ is exit, $y_{i,q,l'} = 0$ is set for all $l' > l$, and the token contributes no further cost to any subsequent layer.

When $H = 1$, the look-ahead term vanishes and the strategy degenerates to a myopic per-layer decision. A larger $H$ produces decisions that are more aware of future transmission costs and remaining budget, at the cost of additional runtime computation. The horizon $H$ therefore provides a tunable tradeoff between decision quality and computational overhead.

*F. Dual-Weight Adaptation*

The weights $\omega_T$ and $\omega_D$ control the tradeoff between inference delay and quality degradation. Since the runtime network condition and server workload vary over time, the weights are periodically updated according to the observed average normalized delay $\bar{T}$ and the observed average budget utilization rate $\bar{D}$ within the latest observation window:

$$\omega_D \leftarrow \Pi_{[0,1]}\left(\omega_D + \kappa(\bar{D} - \bar{T})\right), \omega_T \leftarrow 1 - \omega_D \quad (43)$$

where $\kappa$ is a small positive step size and $\Pi_{[0,1]}(\cdot)$ clips the value into $[0,1]$. The update increases $\omega_D$ when the budget utilization rate is higher than the normalized delay, directing the strategy to be more conservative about quality degradation; conversely, it increases $\omega_T$ when the delay is dominant, directing the strategy toward more aggressive skipping and earlier exit. Since both $\bar{T}$ and $\bar{D}$ are dimensionless quantities normalized in the same way as (21), the update direction is always physically meaningful, and the constraint $\omega_T + \omega_D = 1$ is maintained at every update step. The exit threshold adaptation (37) runs on the same observation window, so the two adaptation rules share a common runtime feedback loop without interfering with each other.

*E. Online Selection Algorithm*

Based on the foregoing models, the communication-aware online skip-exit and server-expert selection algorithm is summarized in Table II.

Table II

Algorithm2: Communication aware online skip exit and server-expert selection

**Input:** Deployment strategy $X$, grouping result $\mathcal{G}$, exit heads $\{h_l\}$, layer importance thresholds $\{\theta_l\}$, transition statistics $\{P_l\}$, look-ahead horizon $H$, quality budgets $\{D_i^{\max}\}$, initial confidence thresholds $\{p_j^i\}$, network and computation parameters.

**Output:** Server-expert selection $U$, skip decisions $\Delta$, exit decisions $Y$, and token trajectories.

1. for each user $d_i$ and each token $q \in Q_i$ do
2. Set initial location $s_{i,q,1} = a(i)$ by (29)
3. Set accumulated degradation $\widehat{D}_{i,q,1} = 0$ and per-token budget $\bar{D}_i$ by (33)
4. end for
5. for each MoE layer $l = 1$ to $N$ do
6. for each user $d_i$ and each token $q \in Q_i$ with $y_{i,q,l} = 1$ do
7. Obtain target expert $r_l(q)$ from the local router
8. Compute layer importance $I_{i,q,l} = g_l^{r_l(q)}(q)$ by (4)
9. Compute transmission necessity score $\Gamma_{i,q,l}$ by (35)
10. Build feasible action set $\mathcal{A}_{i,q,l}$ by (34), admitting skip only if $\Gamma_{i,q,l} > 0$ and $I_{i,q,l} \leq \theta_l$
11. for each action $a \in \mathcal{A}_{i,q,l}$ do
12. Compute per-step cost $f_{i,q,l}(a)$ by (31) or (32)
13. Compute look-ahead cost-to-go $\Phi_{i,q,l}^{(H-1)}(a)$ by the depth-$H$ recursion (39)
14. end for
15. Select $a^* = \arg\min_{a \in \mathcal{A}_{i,q,l}} \left[ f_{i,q,l}(a) + \Phi_{i,q,l}^{(H-1)}(a) \right]$ by (40)
16. if $a^*$ is execute $(m^*, k^*)$ then
17. Set $u_{i,q,l,m^*,k^*} = 1$; execute token $q$ on $S_{m^*}$ by $e_l^{k^*}$
18. Compute confidence $\xi_{i,q,l} = h_l(\mathbf{h}_{i,q,l})$ by (7)
19. Update $s_{i,q,l+1}$ and $\widehat{D}_{i,q,l+1}$ by (41)
20. else if $a^*$ is skip then
21. Set $\delta_{i,q,l} = 1$; update $s_{i,q,l+1}$ and $\widehat{D}_{i,q,l+1}$ by (42)
22. else $\{a^*$ is exit$\}$
23. Set $y_{i,q,l'} = 0$ for all $l' > l$

24. end if
25. end for
26. end for
27. Periodically update $\omega_T$, $\omega_D$ by (43) and $p_J^i$ by (37) from runtime observations
28. return $U$, $\Delta$, $Y$, and token trajectories.

The algorithm contains three parts.

**Lines 1-4: Initialization.** Before the layer-by-layer loop begins, each token's runtime state is initialized. The initial server location $s_{i,q,1}$ is set to the access edge server $a(i)$ of the corresponding user, which is the server to which the user's device is attached and where the token representation first arrives. The accumulated quality degradation $\widehat{D}_{i,q,1}$ is initialized to zero, reflecting the fact that no decision has yet been made. The per-token quality budget $\bar{D}_i$ is derived by evenly distributing the user-level budget $D_i^{\max}$ over all tokens of user $d_i$ according to (33). This even allocation is a sufficient condition for the user-level constraint $D_i \le D_i^{\max}$: as long as each individual token does not exceed its share $\bar{D}_i$, the sum over all tokens is bounded by $D_i^{\max}$, and the constraint is satisfied regardless of the actual exit layers or skip patterns realized at runtime.

**Lines 5-28: Layer-by-layer joint decision.** The outer loop advances from layer $l = 1$ to $N$. Tokens that have already exited ($y_{i,q,l} = 0$) are excluded from processing, so the active set shrinks monotonically as tokens exit, reducing the per-layer computational load as inference progresses. For each active token, the local router is queried to obtain the target expert $r_l(q)$, which is available on the current server without any cross-server communication because the router is replicated on every server following [37]. The layer importance $I_{i,q,l}$ is then computed from the gating weight of the target expert, and the transmission necessity score $\Gamma_{i,q,l}$ is computed by scanning the execute candidates in $\mathcal{A}_{i,q,l}^{\text{exec}}$ for any locally available option. The feasible action set $\mathcal{A}_{i,q,l}$ is then assembled: the execute subset includes all deployed (server, expert) pairs that are substitute-feasible and within the remaining budget; the skip action is admitted only when $\Gamma_{i,q,l} > 0$ and $I_{i,q,l} \le \theta_l$, jointly ensuring that skipping is considered only when execution would incur a genuine transmission cost and the layer importance is low enough that the skip degradation stays within the calibrated tolerance; the exit action is admitted only when the confidence score of the last executed layer already meets the task requirement $p_J^i$, so that exit is never triggered on a representation that has not been sufficiently refined. For each candidate action, the per-step cost is computed, and the look-ahead cost-to-go is evaluated by the depth-$H$ recursion (39), which propagates the expected future transmission cost forward from the server that would result from the current action. The action with the minimum total cost is then selected. In the event that no substitute-execute or skip candidate satisfies the quality budget, the exact-execution candidate with $k^* = r_l(q)$ is selected as the fallback, so that the accumulated degradation never exceeds $\bar{D}_i$ regardless of the routing outcome. The token state is updated according to the chosen action: for execute, both the server location and the accumulated degradation are updated; for skip, only the degradation is updated and the location is inherited unchanged; for exit, the active flags of all subsequent layers are cleared.

**Lines 29-30: Adaptive feedback and output.** After each observation window, the dual weights $\omega_T$ and $\omega_D$ are updated by (43) based on the observed average normalized delay and budget utilization, keeping the online cost function calibrated to the current runtime environment. The exit confidence threshold $p_J^i$ is separately adapted by (37) based on the average exit confidence observed in the same window, so that the threshold self-corrects when the calibration-time and runtime token distributions diverge. The two adaptation rules share the same observation window but operate on disjoint parameters, so they do not interfere with each other. The complete server-expert selection $U$, skip decisions $\Delta$, exit decisions $Y$, and resulting token trajectories are then returned.

*F. Properties of the Proposed Algorithms*

In this section, the feasibility, the quality and confidence guarantees, and the complexity of the two algorithms are analyzed in detail.

**Property 1 (Deployment feasibility)**: If the total edge memory can store every expert at least once, Algorithm 1 outputs a deployment $X$ satisfying constraints (c.1) to (c.3).

*Proof.* We need to show that the output $X$ of Algorithm 1 satisfies constraints (c.1), (c.2), and (c.3).

For (c.3), every deployment variable $x_{m,l,k}$ is set to either 0 or 1 in both the baseline initialization step and the greedy replication loop, so the binary constraint holds throughout.

For (c.2), the baseline deployment from [37] guarantees that every expert $e_l^k$ is placed on at least one server before Algorithm 1's replication loop begins. The replication loop in Lines 12–16 only adds additional copies of existing experts to servers that have residual memory; it never removes an existing replica. Therefore, after the replication loop terminates, every expert is still placed on at least one server, and (c.2) is satisfied.

For (c.1), the baseline deployment from [37] satisfies the memory constraint by assumption. In each iteration of the replication loop, the feasibility check: $p_l^k + \sum_{l',k'} x_{m,l',k'} \cdot p_{l'}^{k'} \le C_m$ is performed before placing any replica. A replica is placed only if this check passes, meaning that the memory footprint on server $S_m$ after placement does not exceed $C_m$. Since the check is enforced at every iteration and the loop only adds replicas that pass the check, the memory constraint on every server is maintained throughout, and (c.1) holds in the final output.

Since (c.1), (c.2), and (c.3) all hold, Algorithm 1 outputs a feasible deployment strategy. ■

**Property 2 (Quality-budget guarantee)**: For any user $d_i$, the strategy produced by Algorithm 2 satisfies $D_i \le D_i^{\max}$.

*Proof.* Fix any user $d_i$ and any token $q \in Q_i$. We show by induction on $l$ that the accumulated degradation $\widehat{D}_{i,q,l}$ satisfies $\widehat{D}_{i,q,l} \le \bar{D}_i$ for every layer $l \in [1, N+1]$.

*Base case.* At initialization, $\widehat{D}_{i,q,1} = 0 \le \bar{D}_i$.

*Inductive step.* Suppose $\widehat{D}_{i,q,l} \le \bar{D}_i$ for some *layer* $l$ at which the token is active. We show $\widehat{D}_{i,q,l+1} \le \bar{D}_i$.

*Execute action* $(m^*, k^*)$**.** The execute *candidate* is admitted into $\mathcal{A}_{i,q,l}$ only if $\widehat{D}_{i,q,l} + Q_l(r_l(q), k^*) \le \bar{D}_i$ by the budget condition in (34). Therefore $\widehat{D}_{i,q,l+1} = \widehat{D}_{i,q,l} + Q_l(r_l(q), k^*) \le \bar{D}_i$.

*Skip action.* The skip action is *admitted* only if $\widehat{D}_{i,q,l} + \Lambda_{i,q,l} \leq \bar{D}_i$. Therefore $\widehat{D}_{i,q,l+1} = \widehat{D}_{i,q,l} + \Lambda_{i,q,l} \leq \bar{D}_i$.

*Exit action.* No further layers are processed, so $\widehat{D}_{i,q,l'}$ remains unchanged for all $l' > l$, and the bound $\widehat{D}_{i,q,l'} \leq \bar{D}_i$ holds trivially.

It remains to verify that at least one feasible action always exists in $\mathcal{A}_{i,q,l}$, so that Algorithm 2 is never forced to violate the budget. By Property 1, every expert is deployed on at least one server, so the exact-execution candidate $(m_0, r_l(q))$ with $k^* = r_l(q)$ exists and satisfies $Q_l(r_l(q), r_l(q)) = 0$ by definition. Therefore, the budget condition reduces to $\widehat{D}_{i,q,l} + 0 \leq \bar{D}_i$, which holds by the inductive hypothesis. Hence the exact-execution candidate is always in $\mathcal{A}^{\text{exec}}_{i,q,l}$, and $\mathcal{A}_{i,q,l}$ is never empty.

By induction, $\widehat{D}_{i,q,N+1} \leq \bar{D}_i$ for *every* token $q$. Summing over all $|Q_i|$ tokens of user $d_i$ and applying the definition $\bar{D}_i = D_i^{\max}/|Q_i|$: $D_i = \sum_{q \in Q_i} \widehat{D}_{i,q,N+1} \leq |Q_i| \cdot \bar{D}_i = D_i^{\max}$.

Therefore, the user-level quality constraint (c.10) is satisfied for every user $d_i$. Crucially, this guarantee holds for any realization of the routing targets and any calibration accuracy of $\theta_l$, because the hard budget check in (34) is performed online at every step independently of any offline approximation. ■

**Property 3 (Confidence-preserving exit)**: Every early exit produced by Algorithm 2 satisfies the confidence requirement, i.e., constraint (c.9) holds.

*Proof.* We need to show that whenever Algorithm 2 sets $y_{i,q,l+1} = 0$ (i.e., token $q$ exits after layer $l$), the confidence constraint $\xi_{i,q,l} \geq p_J^i$ holds, i.e., constraint (c.9) is satisfied.

By the construction of the feasible action set in (34), the exit action is admitted into $\mathcal{A}_{i,q,l}$ if and only if $\xi_{i,q,l-1} \geq p_J^i$ has been observed at the previous executed layer. Consequently, the algorithm can only select the exit action—and thus set $y_{i,q,l+1} = 0$, when this condition is satisfied, and (c.9) holds by construction.

We additionally verify that the confidence score is always evaluated on a representation that has been updated by an executed expert, rather than a stale residual representation from a skipped layer. By constraint (c.6) and its implementation in Algorithm 2, when a layer is skipped ($\delta_{i,q,l} = 1$), the exit action is not admitted into $\mathcal{A}_{i,q,l}$; the active flag $y_{i,q,l+1}$ is passed unchanged from $y_{i,q,l}$ without triggering an exit check. Therefore, $\xi_{i,q,l-1}$ in the exit admission condition always refers to the confidence score produced after the most recent *executed* layer, and the confidence evaluation is never performed on a representation that was not refined by an expert computation.

Finally, by the exit-head training objective (23), the output $\xi_{q,l}$ of exit head $h_l$ approximates the probability that the representation $\mathbf{h}_{q,l}$ is already consistent with the full-depth prediction. Under this calibration, $\xi_{i,q,l} \geq p_J^i$ implies that the representation at layer $l$ agrees with the full-depth output with empirical probability at least $p_J^i$ on the calibration distribution. When the runtime distribution is close to the calibration distribution, this probabilistic guarantee carries over to runtime; when distribution shift occurs, the adaptive threshold update (37) moves $p_J^i$ toward the observed exit confidence, restoring calibration over time. Therefore, every exit decision produced by Algorithm 2 satisfies (c.9), and the exit is quality-preserving by construction. ■

**Property 4 (Complexity)**: Let $|\mathcal{Q}_{\text{cal}}|$ be the calibration set size and $\bar{E}$ the average number of experts per layer. Algorithm 1 runs in $O(N\,|\mathcal{Q}_{\text{cal}}|\,\bar{E})$ time for the single calibration pass and the per-layer statistics, plus $O(M\,N\,\bar{E}\log(MN\bar{E}))$ for the greedy replication. For Algorithm 2, the per-token per-layer decision evaluates at most $|\mathcal{A}_{i,q,l}| \leq M\bar{E} + 2$ actions, each requiring a depth-$H$ look-ahead with branching bounded by the support of the transition distribution $\bar{E}$, so the worst-case per-decision cost is $O((M\bar{E}) \cdot \bar{E}^{H-1})$.

*Proof.* We analyze the complexity of Algorithm 1 and Algorithm 2 separately.

**Algorithm 1.** The dominant cost comes from four steps.

(*i*) *Calibration pass.* Running the frozen model on $|\mathcal{Q}_{\text{cal}}|$ tokens requires a single forward pass. For each of the $N$ MoE layers, extracting the hidden representations and routing targets takes $O(|\mathcal{Q}_{\text{cal}}| \cdot d)$ time, where $d$ is the hidden dimension. Across all layers the total cost is $O(N \cdot |\mathcal{Q}_{\text{cal}}| \cdot d)$.

(*ii*) *Exit-head training.* Each exit head $h_l$ is a single linear layer with $d$ parameters, trained for a fixed number of epochs on $|\mathcal{Q}_{\text{cal}}|$ labels. The training cost per head is $O(|\mathcal{Q}_{\text{cal}}| \cdot d)$, and summing over $N$ layers gives $O(N \cdot |\mathcal{Q}_{\text{cal}}| \cdot d)$. Since $d$ is a model constant, this is $O(N \cdot |\mathcal{Q}_{\text{cal}}|)$ in terms of the problem-size parameters.

(*iii*) *Transition statistics and threshold calibration.* For each layer $l$, computing the $E_l \times E_{l+1}$ transition count matrix requires scanning $|\mathcal{Q}_{\text{cal}}|$ tokens, giving $O(|\mathcal{Q}_{\text{cal}}|)$ per layer and $O(N \cdot |\mathcal{Q}_{\text{cal}}|)$ in total, where $E_l$ is the number of experts at layer $l$. Computing the skip degradation curve and deriving $\theta_l$ also requires $O(|\mathcal{Q}_{\text{cal}}|)$ per layer. The centroid clustering with a fixed number of centroids $K_c$ runs in $O(|\mathcal{Q}_{\text{cal}}| \cdot K_c)$ per layer.

(*iv*) *Greedy replication.* Each replication iteration evaluates $B_{m,l,k}$ for at most $M \cdot N \cdot \bar{E}$ (server, expert) pairs, where $\bar{E}$ is the average number of experts per layer and $M$ is the number of servers. With at most $M \cdot N \cdot \bar{E}$ total replication decisions, the replication loop runs in $O(M^2 N \bar{E}^2)$ time in the worst case, or $O(MN\bar{E}\log(MN\bar{E}))$ if the benefit values are maintained in a max-heap.

Combining all steps, the total complexity of Algorithm 1 is $O(N \cdot |\mathcal{Q}_{\text{cal}}| \cdot d + MN\bar{E}\log(MN\bar{E}))$. Since both terms are polynomial in the problem parameters, Algorithm 1 has polynomial-time complexity.

**Algorithm 2.** For each active token $q$ at each layer $l$, the feasible action set $\mathcal{A}_{i,q,l}$ contains at most $M\bar{E} + 2$ actions (at most $M\bar{E}$ execute candidates, plus skip and exit). For each action, the depth-$H$ look-ahead recursion (39) is evaluated. At each recursion level, the sum over $k'$ in (39) has at most $\bar{E}$ non-zero terms (the support of the transition distribution), and the inner minimization is over $M\bar{E} + 2$ candidates. Therefore the cost of one complete look-ahead evaluation is bounded by: $\underbrace{(M\bar{E} + 2)}_{\text{branching at } l} \cdot \underbrace{\bar{E}^{H-1}}_{\text{subsequent levels}} \cdot \underbrace{(M\bar{E} + 2)}_{\text{inner min per node}} = ((M\bar{E})^2 \cdot \bar{E}^{H-1})$.

Summing over all tokens and layers, the total cost of Algorithm 2 is $O(N \cdot |Q| \cdot (M\bar{E})^2 \cdot \bar{E}^{H-1})$, where $|Q|$ is the

total number of tokens across all users. This is polynomial in all problem parameters for any fixed horizon $H$.

In practice, the transition distribution $P_l(k \to k')$ is sparse because each token is routed to a small number of experts per layer, so the effective branching factor at each look-ahead level is much smaller than $\bar{E}$. When $H = 1$ the look-ahead term vanishes entirely and the per-token per-layer cost reduces to $O(M\bar{E})$, which is the same order as a simple myopic selection. Therefore, the horizon $H$ controls the complexity-quality tradeoff continuously, and the algorithm remains practically efficient for the small values of $H$ used in the experiments. ■

## VI. Performance Evaluation

In this section, we evaluate the proposed TrimMoE framework by experiment and analyze its behavior from the perspectives of latency, communication, depth reduction, and inference quality.*A. Experimental Setup*

**Experimental setup.** TrimMoE is implemented on PyTorch and deployed on a geo-distributed edge testbed of 10 physical edge servers. The servers are heterogeneous in both computation and memory: their computation capacities $F_m$ span 18 to 120 TFLOPS and their GPU memory capacities $C_m$ span 12 to 48 GB. The servers are interconnected over ordinary Internet links rather than a dedicated fabric, with available bandwidths $B_{m,n}$ ranging from 0.8 to 10 Gbps and propagation-plus-queuing latencies $\tau_{m,n}$ ranging from 0.3 to 22 ms, so that the bandwidth-limited and time-varying nature of edge interconnections is reproduced. Each user is attached to one access edge server following a non-uniform spatial distribution, and requests arrive at a default rate of 40 requests per second unless stated otherwise.

**Models, tasks, and calibration.** We evaluate TrimMoE on three MoE models of increasing scale, Switch-Base-8E, Qwen-MoE-A2.7B, and Mixtral-8x7B. Inference quality is measured on WikiText-103 (perplexity), SQuAD (F1), and GSM8K (accuracy). All offline quantities required by Algorithm 1 (the per-layer exit heads, the layer-importance thresholds, the expert transition statistics, and the token-similarity reference set) are computed once on a held-out calibration set with the backbone strictly frozen. Unless otherwise stated, the look-ahead horizon is H=3, the per-user quality-degradation budget is set to 2%, the confidence requirement is $p_j^i = 0.9$, and the memory budget ratio of the redundant deployment is 2.0.

**Baselines.** We compare TrimMoE with five representative methods that cover the relevant design space: 1) EdgeShard [12], 2) SlimCaching [30], 3) MoE² [35], 4) Mixture-of-Depths (MoD) [21], and 5) Substitute-Only [37], our previous similarity-based substitute-execution framework without depth reduction. We further include four internal variants, i.e., TrimMoE without skip, without early exit, without look-ahead, and without communication-aware redundancy, for the ablation study. The metrics are the average inference latency, the tail (P99) latency, the cross-server communication volume, the remote execution ratio, the throughput, and the inference quality.

### B. Overall Performance Comparison

We first compare the overall performance on Mixtral-8x7B, as shown in Fig. 4, which reports the average latency, the cross-server traffic per one thousand tokens, the remote execution ratio, and the throughput.

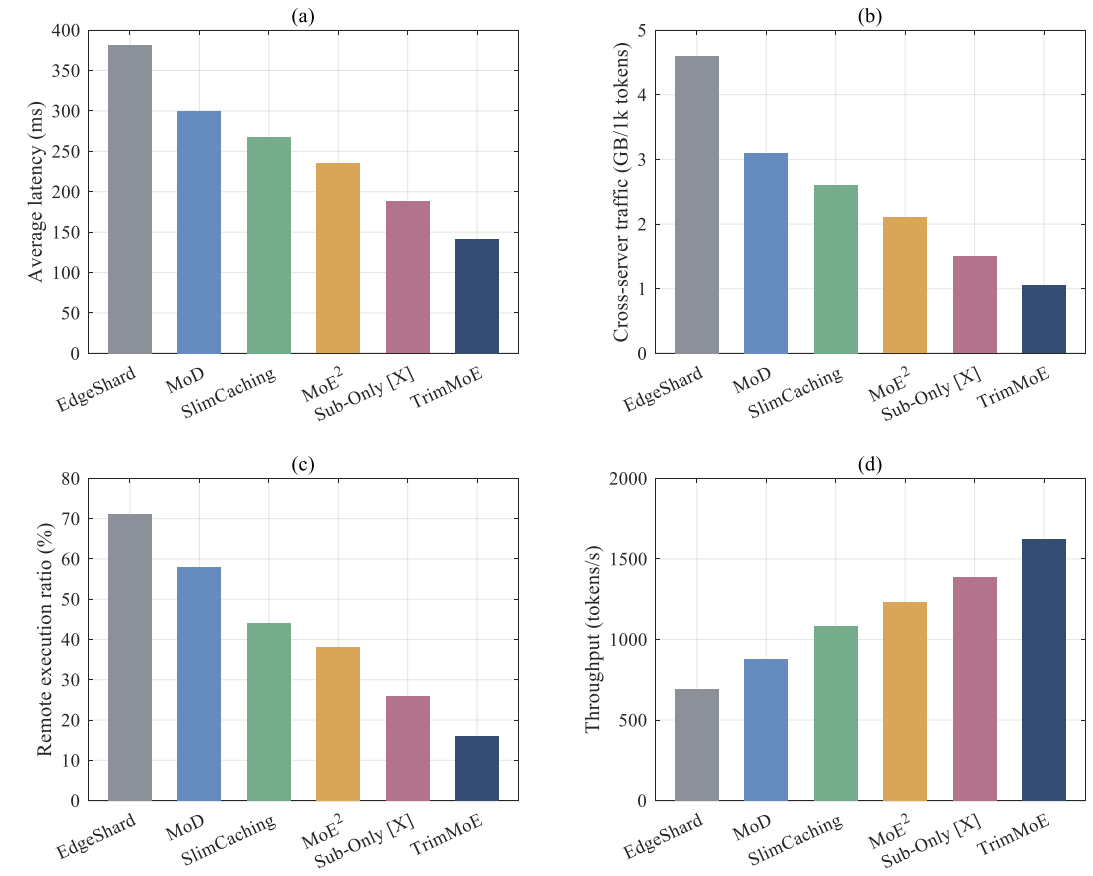

Fig. 4 Overall performance

As shown in Fig. 4(a), TrimMoE attains the lowest average latency of 142ms, which is 62.8% lower than EdgeShard, 39.6% lower than MoE², and 24.5% lower than our prior Substitute-Only scheme. The additional gain over Substitute-Only is the direct effect of the two depth-reduction mechanisms: a low-importance layer that would otherwise require a remote hop is skipped on the spot, and once the confidence requirement is met the remaining layers, together with their potential cross-server transmissions, are terminated at once. Fig. 4(b) shows that the cross-server traffic of TrimMoE is only 1.05 GB per one thousand tokens, 77.2% below EdgeShard and 30.0% below Substitute-Only, and Fig. 4(c) shows that its remote execution ratio drops to 16%, far below the 71% of EdgeShard and the 26% of Substitute-Only. Consistently, Fig. 4(d) shows that TrimMoE delivers the highest throughput of 1620 tokens per second, 2.35× that of EdgeShard. It is worth noting that MoD reduces latency relative to EdgeShard but remains clearly worse than the MoE-aware schemes, because it skips layers without being aware of which executions are cross-server, so it often drops cheap local layers while keeping the expensive remote ones.

### C. Tail Latency and Execution Behavior

Fig. 5 reports the latency distribution and the expert execution-type breakdown. Fig. 5(a) plots the empirical CDF of the per-request latency, and Fig. 5(b) decomposes all MoE layer-level executions into local exact, local substitute, remote exact, remote substitute, skipped, and early-exited.

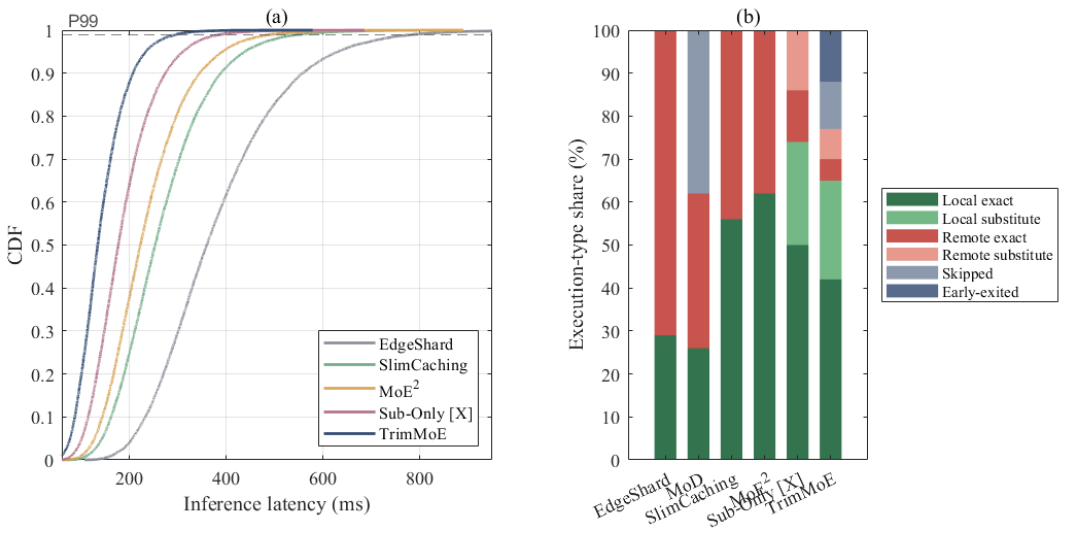

Fig. 5 Tail latency and execution behavior

From Fig. 5(a), TrimMoE not only shifts the whole distribution to the left but also compresses the tail: its P99 latency is 268ms, 41.1% lower than that of MoE² and 64.0% lower than that of EdgeShard. The compressed tail comes from the look-ahead selection, which avoids moving a token onto a server that is favorable now but expensive for the subsequent layers, thereby eliminating the worst-case multi-hop trajectories. From Fig. 5(b), TrimMoE removes 23% of the layer executions entirely, 11% skipped and 12% terminated by early exit, and among the remaining executions only 16% are remote. In contrast, EdgeShard executes every layer exactly and incurs 71% remote executions, while Substitute-Only still runs to full depth and leaves 26% remote executions. This confirms that TrimMoE attacks the two residual sources of cross-server traffic that substitute execution alone cannot remove.

### *D. Inference Quality and the Quality Budget*

Fig. 6 evaluates the inference quality. Fig. 6(a) reports the quality of representative methods on the three datasets, and Fig. 6(b) shows how the latency and the actual quality degradation of TrimMoE vary with the per-user quality budget.

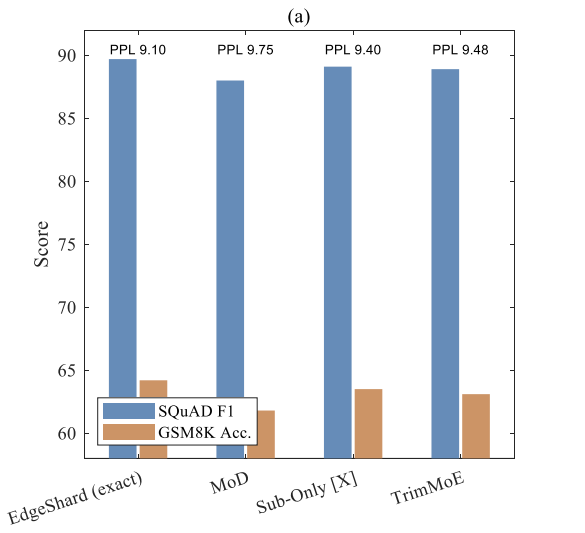

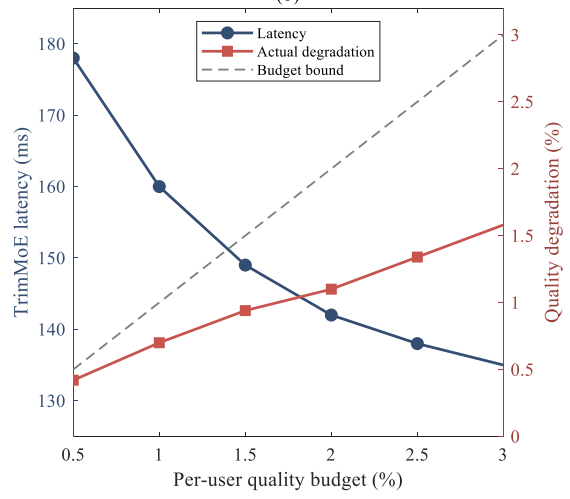

Fig. 6 Inference quality and quality budget

From Fig. 6(a), the quality of TrimMoE stays close to the exact methods: compared with the full centralized model, the perplexity on WikiText-103 rises by only 0.38, the SQuAD F1 drops by only 0.8, and the GSM8K accuracy drops by only 1.1. This small and controlled loss is a consequence of the design: early exit is gated by the confidence condition and is therefore quality-preserving by construction, while skipping is admitted only for low-importance layers within the offline-calibrated tolerance and the online per-token budget. By comparison, MoD skips layers without a confidence guarantee and suffers a larger GSM8K drop of 2.4. From Fig. 6(b), as the quality budget is relaxed from 0.5% to 3%, TrimMoE trades quality for latency in a smooth and monotone manner, and the measured degradation always stays strictly below the budget bound, which is exactly the hard guarantee established in Property 2. The knee of the curve is around a 2% budget, where the latency is already close to its minimum while the degradation is still well controlled, which justifies the default setting.

### *E. Depth-Reduction Behavior*

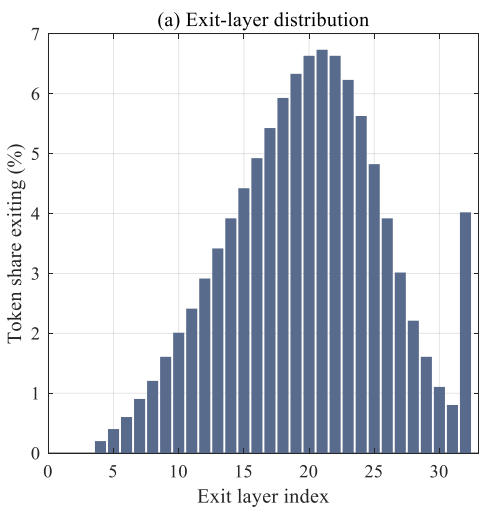

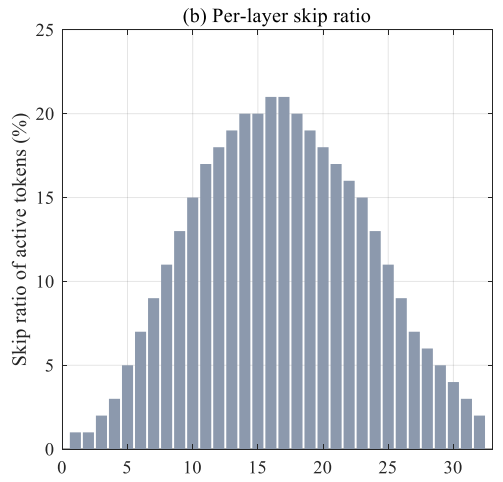

Fig. 7 Depth reduction behavior

To examine how TrimMoE allocates depth, Fig. 7 reports the exit-layer distribution and the per-layer skip ratio on Mixtral-8x7B. From Fig. 7(a), most tokens are confidently terminated in the deep layers, with the distribution peaking around layers 20 to 22 and a residual mass at the final layer for the hardest tokens, which indicates that a large fraction of tokens do not need the full depth once their representation has stabilized. From Fig. 7(b), the skip ratio is low in the shallow layers, where experts are more specialized and important, and rises in the middle-to-deep layers, where the routed expert contributes marginally and can be bypassed through the residual path. This depth-aware behavior is consistent with the layer-importance model and explains why the quality loss remains bounded even though nearly a quarter of the layer executions are eliminated.

### *F. Effect of the Look-Ahead Horizon*

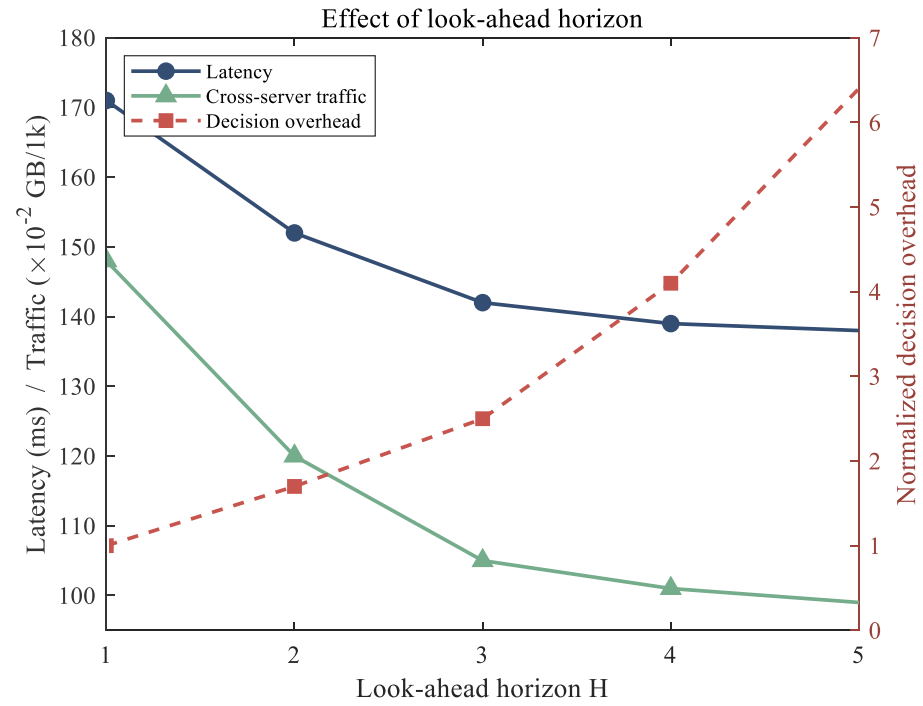

Fig. 8 Effect of the look-ahead horizon

Fig. 8 evaluates the effect of the look-ahead horizon H on Mixtral-8x7B, with H varied from 1 to 5, where H=1 reduces to the myopic per-layer decision. As H grows from 1 to 3, both the average latency and the cross-server traffic decrease quickly, by 17.0% and 29.1% respectively, because anticipating more future layers prevents the strategy from skipping or relocating a token in a way that merely defers an expensive transmission to a later hop. Beyond H=3 the two curves saturate, while the per-token decision overhead keeps growing because the cost-to-go recursion expands more candidate trajectories. Thus H=3 offers a good balance between decision quality and runtime overhead, in agreement with the complexity analysis in Property 4.

### *G. Comparison Across MoE Models*

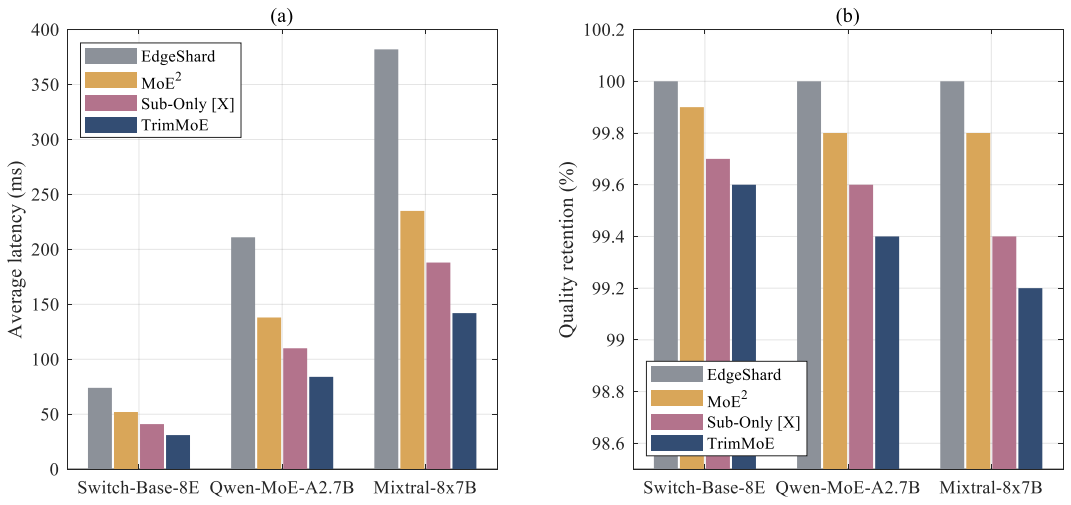

Fig. 9 Comparison across MoE models

Fig. 9 compares TrimMoE with the representative baselines on the three MoE models. As shown in Fig. 9(a), TrimMoE consistently achieves the lowest latency, and its advantage widens as the model scales: the latency reduction relative to MoE² grows from 40.4% on Switch-Base-8E to 39.6% on Mixtral-8x7B, while the absolute saving grows from 21ms to 93 ms. Larger models have more layers and more experts, which both increases the number of cross-server-prone layers that skipping and early exit can remove and enlarges the pool of local substitutes. As shown in Fig. 9(b), the quality retention of TrimMoE stays above 99.2% on all three models, confirming that the latency gains do not come at the cost of inference quality.

### *H. Ablation Study*

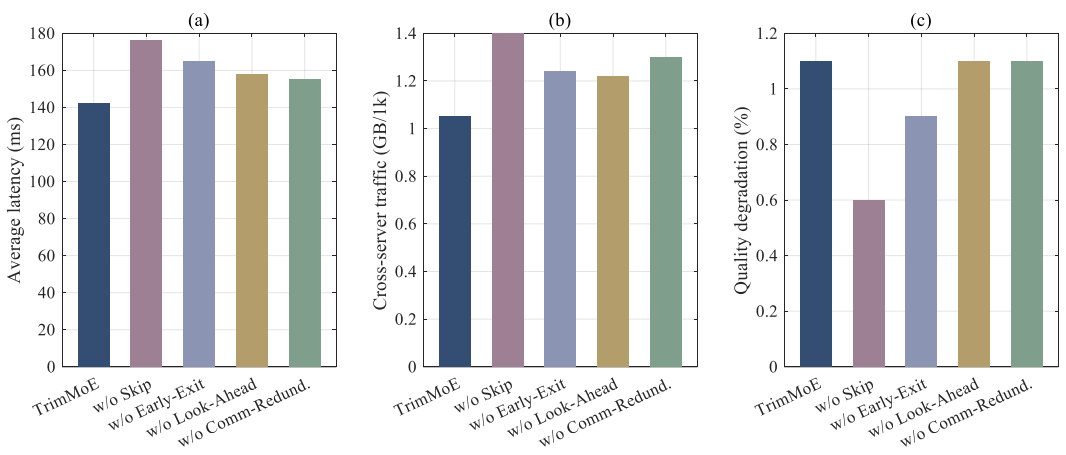

Fig. 10 Ablation Study

Fig. 10 isolates the contribution of each component on Mixtral-8x7B by removing the skip mechanism, the early exit, the look-ahead, and the communication-aware redundancy in turn. From Fig. 10(a) and Fig. 10(b), removing any component increases both latency and cross-server traffic. Disabling the skip mechanism causes the largest latency increase, from 142ms to 176ms, because the token must then traverse every layer and pay every cross-server hop that an importance-based skip would have removed; disabling early exit and look-ahead also degrades performance, though to a smaller extent. From Fig. 10(c), the quality degradation of all variants stays small and even decreases slightly when skipping is removed, which shows that the depth-reduction components mainly serve to cut latency and communication rather than to preserve quality—the quality is already protected by the confidence gate and the per-user budget regardless of which component is enabled.

### *I. Performance Under Different Request Loads*

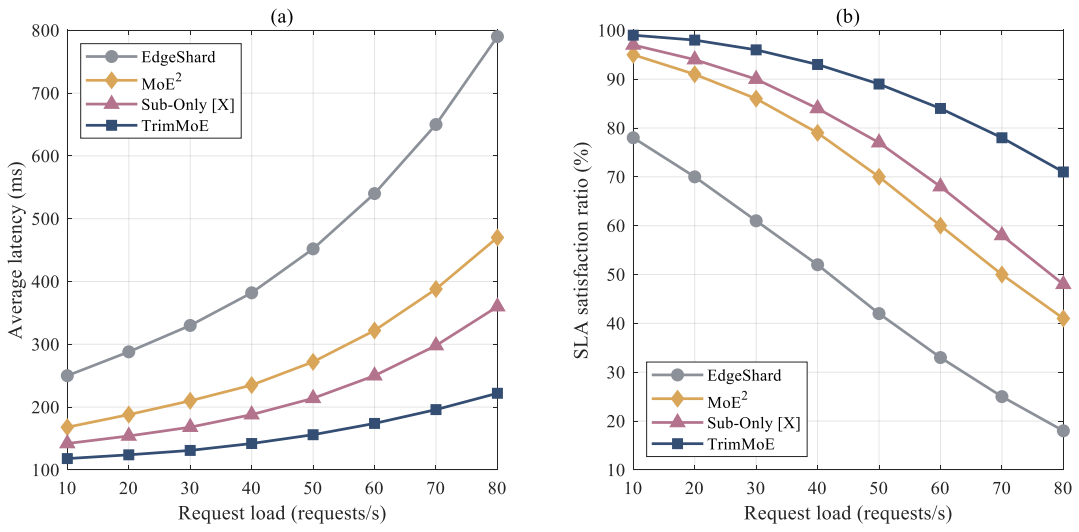

Fig. 11 Performance Under Different Request Loads

Finally, Fig. 11 evaluates robustness to the request load, varied from 10 to 80 requests per second. As shown in Fig. 11(a), the average latency of all methods grows with the load because of increased GPU queueing and link contention, but TrimMoE grows the slowest, since fewer and shorter cross-server transmissions relieve both the busy links and the GPU queues. As shown in Fig. 11(b), where the SLA satisfaction ratio is defined as the fraction of requests completed below a 300ms target, TrimMoE still satisfies 71% of requests at 80 requests per second, whereas the baselines fall below 50% under the same load. This confirms that, by removing unnecessary layer executions and the transmissions they trigger, TrimMoE remains markedly more robust under heavy load.

## VII. Conclusion

This paper reframed distributed-edge MoE serving from accelerating remote-expert access to eliminating unnecessary layer executions and the transmissions they trigger. TrimMoE jointly optimizes communication-aware layer skipping, confidence-based early exit, substitute execution, and server-expert selection under a quality budget, with a transition-aware look-ahead and a skip/exit-aware expert replication. Across three MoE models on a heterogeneous 10-server testbed, TrimMoE substantially lowers average and tail latency as well as cross-server traffic while preserving task quality and remaining robust under heavy load. Our analysis delineates the guarantee scope clearly: the substitution-and-skipping proxy degradation is hard-bounded, while exit quality is governed by the confidence gate and adaptive calibration. Future work includes skip-conditioned transition modeling to tighten look-ahead accuracy and extending the attention-placement model to settings where attention is sharded across servers.

## Acknowledgment

This work was supported in part by the grant from NSFC Grant no. 62571156, 62101159, 52475009, NSF of Shandong Grant no. ZR2021MF055, ZR2025QC666, the Research Grants Council of Hong Kong under the Areas of Excellence scheme grant AoE/E-601/22-R, and also the Opening Project of the Key Laboratory of Advanced Manufacturing and Intelligent Technology (Ministry of Education) at Harbin University of Science and Technology (KFKT202306).

Additionally, the authors used an AI-based language assistance tool to improve the clarity and readability of parts of the manuscript, particularly the Abstract and Introduction, and all technical content, analysis, and conclusions were developed and carefully verified by the authors.